\documentclass[aps,prb,superscriptaddress,showpacs,reprint,floatfix]{revtex4-1}
\usepackage{bm}
\usepackage[dvips]{graphicx}
\usepackage[utf8]{inputenc}
\usepackage[T1]{fontenc}

\begin{document}

\title{Proton configurations in the hydrogen bonds of KH$_2$PO$_4$ as seen by resonant x-ray diffraction}
\author{G. Beutier}
\email{guillaume.beutier@simap.grenoble-inp.fr}
\affiliation{CNRS, SIMAP, F-38000 Grenoble, France}
\affiliation{Univ. Grenoble Alpes, SIMAP, F-38000 Grenoble, France}
\author{S. P. Collins}
\author{G. Nisbet}
\affiliation{Diamond Light Source, Harwell Science \& Innovation Campus, Didcot, Oxfordshire OX11 0DE, United Kingdom}
\author{K. A. Akimova}
\author{E. N. Ovchinnikova}
\author{A. P. Oreshko}
\affiliation{M. V. Lomonosov Moscow State University, Faculty of Physics, 119991 Moscow, Russia}
\author{V. E. Dmitrienko}
\affiliation{A.V. Shubnikov Institute of Crystallography, Russian Academy of Sciences, 119333, Moscow, Russia}

\begin{abstract}

KH$_2$PO$_4$ (KDP) belongs to the class of hydrogen-bonded
ferroelectrics, whose paraelectric to ferroelectric phase
transition is driven by the ordering of the protons in the
hydrogen bonds. We demonstrate that forbidden reflections of KDP,
when measured at an x-ray absorption edge, are highly sensitive to the
asymmetry of proton configurations. The change of average symmetry caused by
the 'freezing' of the protons during the phase transition is
clearly evidenced. 
In the paraelectric phase, we identify in the resonant spectra of the forbidden reflections 
a contribution related to the transient
proton configurations in the hydrogen bonds, which violates the
high average symmetry of the sites of the resonant atoms.
The analysis of the temperature dependence reveals a change
of relative probabilities of the different proton configurations.
They follow the Arrhenius law and the activation energies of polar and Slater configurations are 18.6 and 7.3 meV respectively.

\end{abstract}


\maketitle

\section{Introduction}

\begin{figure*}
    \includegraphics[width=0.5\columnwidth]{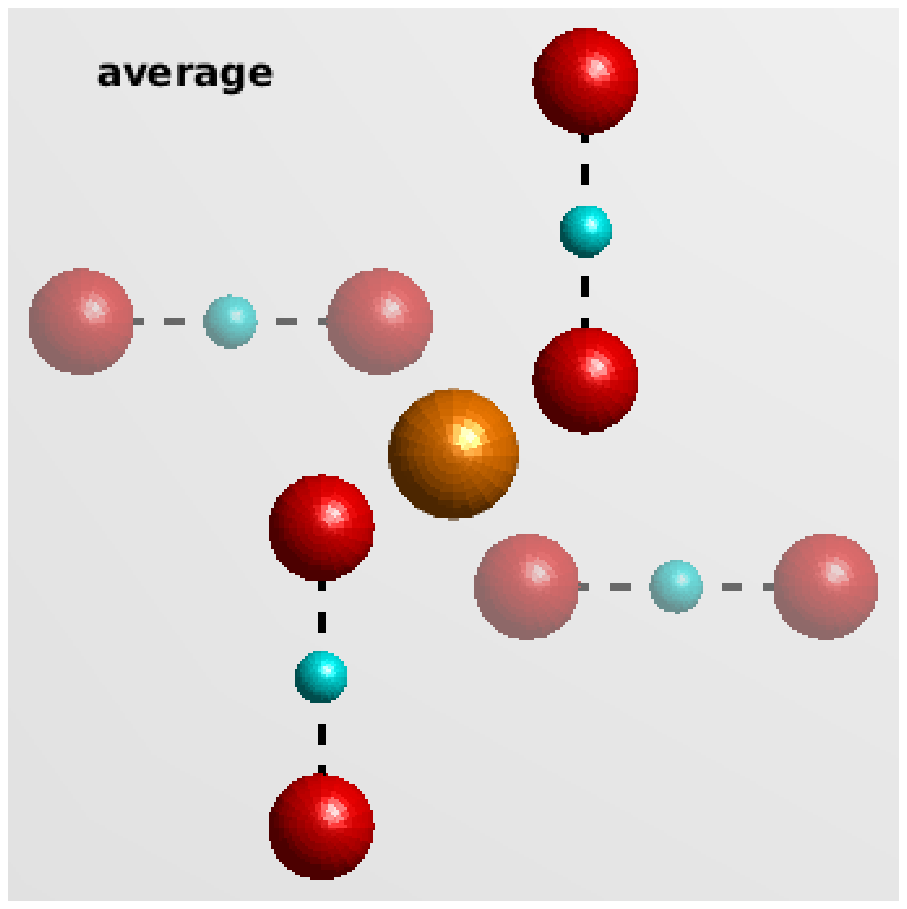}
    \includegraphics[width=0.5\columnwidth]{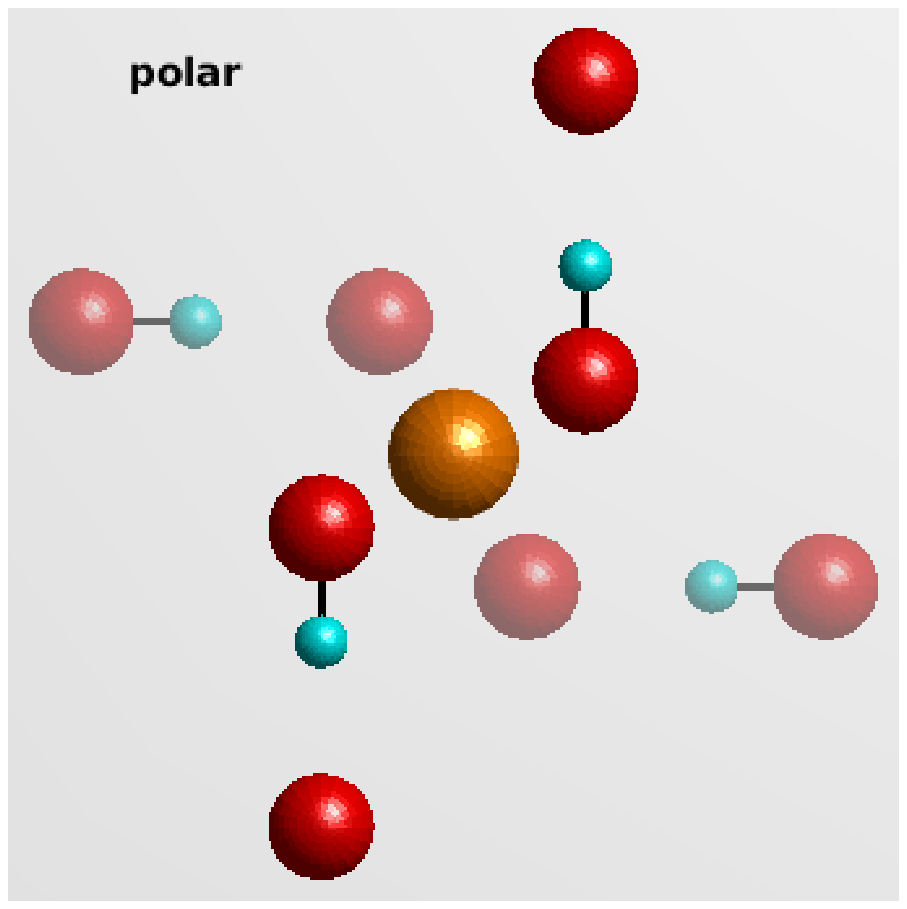}
    \includegraphics[width=0.5\columnwidth]{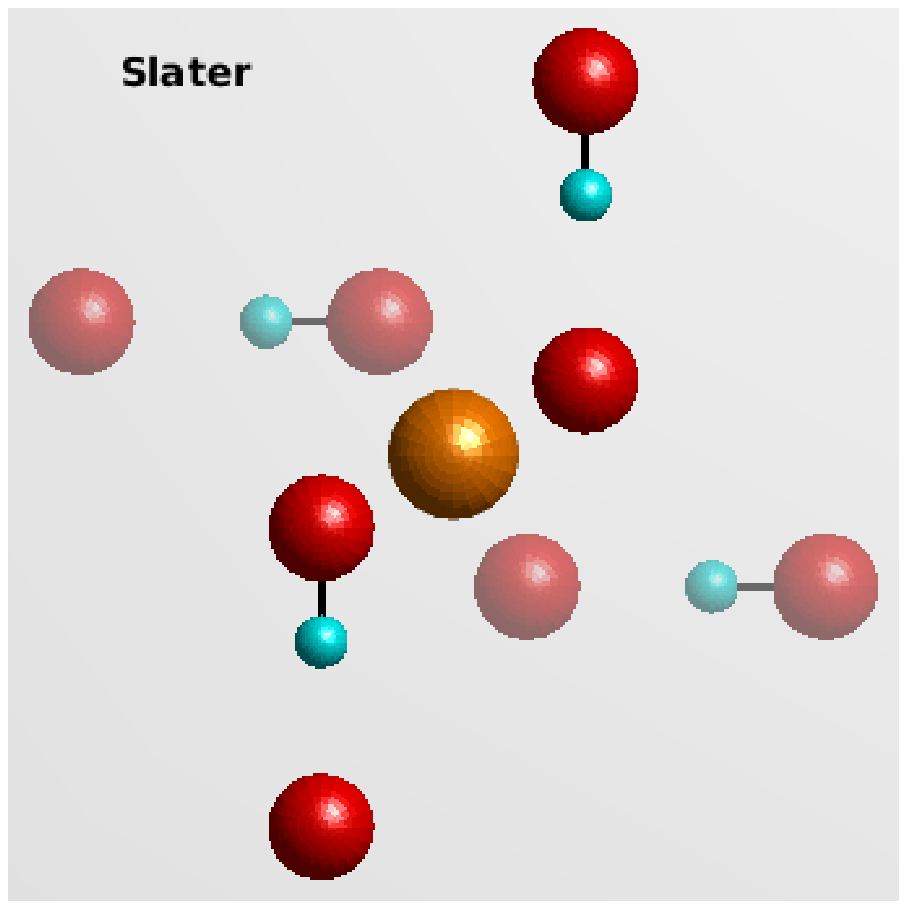}
    \includegraphics[width=0.5\columnwidth]{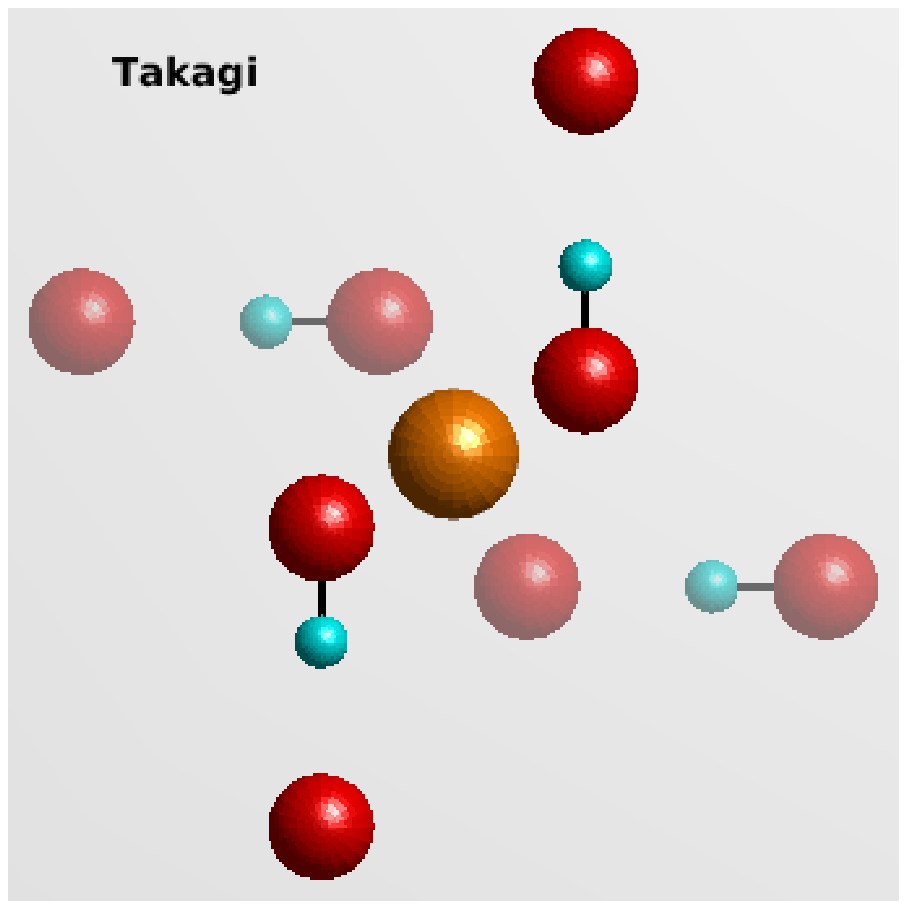}
    \caption{Possible proton configurations in the four H bonds of a PO$_4$ group. 
    The central phosphorus atom is shown in yellow, oxygen atoms are in red, and hydrogen atoms are in light blue. 
    The potassium atoms, situated above and below the phosphorus atoms, are omitted for clarity. 
    The shading effect denotes a different height perpendicular to the plane of the figure.
    The average configuration is fully symmetrical and does not allow E1E1 scattering at the forbidden reflections, but the real configurations do.
    }
    \label{fig.Hconfigs}
\end{figure*}

Although potassium dihydrogenphosphate (KH$_2$PO$_4$, hereafter KDP) 
was one of the first discovered ferroelectric materials \cite{iona}, 
the microscopic mechanism at play during its ferroelectric phase 
transition has been one of the most difficult to understand. 
The crystals of the KDP family belong to the class of hydrogen-bonded 
ferroelectrics, in which protons play an important role: 
their PO$_4$ molecular units are linked by hydrogen bonds, 
and ferroelectricity appears to be connected to the behaviour 
of the protons in these bonds. 
The generic theoretical framework describing the hydrogen-bonded ferroelectrics was introduced by Slater \cite{slater}: 
the static and dynamic properties of these systems are described on the basis of the configuration energy
determined by proton configurations (Fig. \ref{fig.Hconfigs}). 
In Slater's model, each proton occupies one of two possible crystallographic positions in its bond. 
In the paraelectric phase, both positions are equivalent and randomly occupied, while in the ferroelectric phase one of the positions is favoured, according to the local ferroelectric polarisation. 
The ferroelectric transition appears thus as a classical order-disorder phase transition \cite{slater,takagi1,senko}.
Intensive experimental and theoretical investigation has confirmed this model \cite{strukov74}. 
The proton ordering at the phase transition has been evidenced and correlated with atomic displacements along the $c$-axis at the origin of the electric polarisation \cite{nelmes1985}.
To explain the large effect of deuteration on the transition, Blinc suggested that, instead of a static proton disorder, 
protons are in fact delocalised and tunnel back and forth between both sites of a double-well potential \cite{blinc}.
Since then, the nature of the phase transition, either order-disorder or confinement-deconfinement, has been much debated 
(see, \textit{e.g.}, reviews by Schmidt \cite{schmidt}, Tokunaga \& Matsubara \cite{tokunaga} and Lines \& Glass \cite{lines}).
Geometrical effects have been suggested as an alternative to tunnelling to explain the modification of the phase transition of deuteration \cite{ichikawa1981,ichikawa1987}. 
Nowadays there is growing evidence for a combination of both effects \cite{nelmes1988,bussmannholder}. 
Indeed, recent \textit{ab initio} calculations \cite{koval2002,koval2005} confirm the interplay of geometrical and tunneling effects, 
and provide a more accurate view of the tunneling mechanism, which involves not only protons but larger clusters including heavy atoms.
Experimentally, recent neutron Compton scattering experiments have shown strong evidence for the deconfinement of the protons in the paraelectric phase \cite{reiter2002}.
While the role of the protons during the phase transition is now well established, 
not much attention has been given to their behaviour in the paraelectric phase.
In particular, a number of different proton configurations have been proposed \cite{slater,takagi2} 
and are expected to coexist with different probabilities, but a quantitative experimental 
evaluation of their probabilities is still lacking.

In the present paper, we report on a spectroscopic study of the forbidden reflections of KDP with resonant x rays. 
We recently demonstrated that such forbidden reflections show spectacular 
effects across the phase transition \cite{beutier14}.
Here we go further by carefully modeling the spectra and their 
temperature dependence in the paraelectric phase; 
we determine the relative probabilities of various proton configurations 
and show that they change with temperature.
Recently, a similar effect has been studied in rubidium dihydrogen phosphate (hereafter RDP), 
whose structure is isomorphic to that of KDP \cite{richter14}. Due to the limited number 
of data sets and the limited temperature range, only one particular 
proton configuration (the so-called Slater configuration - see below) 
was evidenced. Here, by studying two different types of forbidden reflections 
(the $00l$ and $hhh$ reflections) at two different azimuths and over a large 
temperature range, we are able to extract the contributions of all three main proton configurations.
To this purpose, we present a methodology to deal quantitatively 
with the influence of defects in resonant elastic x-ray scattering (REXS), 
the scope of which goes well beyond the particular case of KDP.

\section{Forbidden reflections in the para- and ferroelectric phases
\label{theory}}

\subsection{General structure factor for the forbidden reflections}

The paraelectric-ferroelectric phase transition of KDP occurs at
the Curie temperature $T_c=123$ K. The paraelectric and
ferroelectric phases have, respectively, body-centered tetragonal ($I\bar 42d$,
$Z=4$) and face-centered orthorhombic ($Fdd2$, $Z=8$) systems,
and ferroelectricity appears along the $c$-axis. 
In the paraelectric (tetragonal) phase of KDP, each proton 
tunnels back and forth between 
two sites of equal probabilities related by symmetry. The
disordered distribution of each proton between two oxygen atoms of
the hydrogen bond has been confirmed by neutron diffraction \cite{nelmes1982}. 
The description of the paraelectric phase by the $I\bar 42d$ group
corresponds to the structure averaged over the proton distribution.

The usual settings used for the description of the ferro- and
paraelectric phases differ by a rotation of 45$^\circ$ around the
$c$ axis accompanying a doubling of the unit cell. Below we shall
use the settings which correspond to the paraelectric (disordered) phase 
\footnote{unlike in Ref. \onlinecite{beutier14}}. 
In these settings, Bragg reflections with Miller indices $hhl$ 
such that $2h+l=4n+2$ are forbidden in conventional 
x-ray diffraction in both phases, due to a glide-plane symmetry. 
However they appear with significant intensity when the energy 
of the incoming x rays is tuned close to an absorption edge, 
due to the anisotropy of the tensor of scattering \cite{dmitrienko83}: 
indeed we recently reported the observation of the $002$ and $222$ 
reflections at the potassium K edge \cite{beutier14}. 
We showed that the intensity and the energy spectra of these reflections undergo huge changes across the phase transition, 
because the electric dipole-dipole (E1E1) resonant scattering vanishes in the higher symmetry (tetragonal) phase.

The general structure factor $F$ of the Bragg reflections 
with Miller indices $hhl$, $2h+l=4n+2$, is equal to
\begin{equation}
 F =  2\left( f^{(1)} - f^{(2)} \right)
\label{F}
\end{equation}
where  $f^{(1)}$ and  $f^{(2)}$ are the atomic scattering factors
of two potassium atoms related by a glide plane symmetry, for instance those
with coordinates $(0 0 \frac{1}{2})$ and $(\frac{1}{2} 0 \frac{1}{4})$. 
Both atoms have essentially the same atomic scattering factors off-resonance, 
but they become highly sensitive to the local anisotropy 
when the incident x rays excite one of their electronic transitions, 
providing a sizable difference of atomic scattering factors, 
which in turn allows for the existence of these pure resonant forbidden reflections.

\subsection{Resonant elastic x-ray scattering}

REXS is usually described by a series of electric multipoles. 
In the following, Cartesian tensors will be used to describe the x-ray polarisation dependence 
of the atomic scattering factors and of their structure factors.
The atomic scattering factor, expanded up to the quadrupolar terms, can be written \cite{blume}:
\begin{equation}
 f = \epsilon_\alpha^{\prime\ast} \epsilon_\beta 
    \left[ f^{dd}_{\alpha\beta} 
    + \frac{i}{2} \left( k_\gamma f^{dq}_{\alpha\beta\gamma} - k^\prime_\gamma f^{dq\ast}_{\beta\alpha\gamma} \right)
    + \frac{1}{4} k_\gamma^\prime k_\delta f^{qq}_{\alpha\gamma\beta\delta}
    \right]
 \label{f}
\end{equation}
with the implicit sum over the indices $\alpha,\beta,\gamma,\delta\in\{x,y,z\}$.
The tensors $\bm f^{dd}$, $\bm f^{dq}$, and $\bm f^{qq}$ stand for the 
electric dipole-dipole (E1E1), dipole-quadrupole (E1E2), and quadrupole-quadrupole (E2E2) resonances, respectively.
$\bm k$ and $\bm k^\prime$, on one hand, and $\bm \epsilon$ and $\bm \epsilon^\prime$, on the other hand, 
are the incident and scattered wave vectors and polarisation states.
In the following we will use the usual decomposition of the x-ray polarisation onto the basis vectors $\sigma$ and $\pi$, 
respectively perpendicular and parallel to the scattering plane.
We shall also use $\bm H = \bm k^\prime - \bm k$ and $\bm L = \bm k^\prime + \bm k$.

The E1E1 term, described by the second-rank tensor $\bm f^{dd}$, is usually largely dominant.
This is the case in the ferroelectric phase, in which the twofold axial symmetry of the resonant site
allows for a non vanishing tensor component ($f^{dd}_{xy}$) in the structure factor.
But, in the paraelectric phase, the twofold axis turns into a pseudofourfold axis (symmetry $\bar 4$), 
cancelling all off-diagonal elements of symmetric second-rank tensors such as $\bm f^{dd}$:
the glide-plane extinction rule still applies \cite{beutier14}.

In the absence of E1E1 contribution, weaker terms become important.
The E1E2 and E2E2 terms, described by the third-rank tensor $\bm f^{dq}$
and the fourth-rank tensor $\bm f^{qq}$ respectively, are the most obvious candidates, 
and a symmetry analysis shows that they actually do not vanish at the forbidden reflections considered here.
We will nevertheless ignore the E2E2 term, which is believed to be much weaker than the E1E2 term in this case, 
based on spectroscopic calculations with the code FDMNES \cite{bunau,fdmnes}, 
and cannot account for the temperature dependence reported below.

The E1E2 term (as well as the E2E2 term) is essentially temperature-independent, 
despite the small variation of the crystal structure in absence of phase transition \cite{oreshko12}, 
and cannot account for the temperature dependence of forbidden reflections in Ge, ZnO and GaN, 
in which the E1E1 term also vanishes \cite{kokubun2001,kirfel2002,collins03,beutierGaN}: 
in these systems, the intensity of forbidden reflections increases with temperature, 
despite the Debye-Waller effect, and the intensity growth is accompanied by a change of spectrum 
that can only be explained by interference with a second scattering process \cite{collins03,beutierGaN}. 
The latter was ascribed to thermal-motion induced (TMI) scattering \cite{dmitrienko2000}.
This mechanism is also expected in KDP.
Similarly to what has been done for Ge, ZnO, and GaN, we will
assume that the main contribution to the TMI term comes from the
displacement of the resonant ion. 
The TMI structure factor can be written \cite{dmitrienko2000}:
\begin{equation}
F_{\alpha\beta}^{TMI}= i H_\delta \frac{\partial f_{\alpha\beta}}{\partial u_\gamma} \langle u_\gamma u_\delta \rangle 
 \equiv i H_\gamma \frac{\partial f_{\alpha\beta}}{\partial u_\gamma} \langle u_\gamma^2\rangle
\label{tmi}
\end{equation}
where $\bm u$ is the displacement of the resonant atom (potassium in our case), 
and the implicit sum over indices $\gamma$ and $\delta$ is assumed. 
The right part of (\ref{tmi}) is valid for (at least) orthorhombic point symmetry, which is the case here.
The mean-square components $\langle u_\gamma^2\rangle$ provide the temperature dependence of this term. 
We define for the following the TMI third-rank tensor $f^{TMI}_{\alpha\beta\gamma}$ as:
\begin{equation}
 f_{\alpha\beta\gamma}^{TMI} = i \frac{\partial f_{\alpha\beta}}{\partial u_\gamma} \langle u_\gamma^2\rangle
\end{equation}
such that $f^{TMI}_{\alpha\beta\gamma}$ is intrinsic to the material 
and couples with the beam according to $F_{\alpha\beta}^{TMI} = H_\gamma f_{\alpha\beta\gamma}^{TMI}$.

It will be shown below that the E1E2 and TMI terms alone cannot explain the experimental results in KDP. 
We need to consider an additional contribution to the resonant atomic factor, 
which is provided by transient proton configurations \cite{mukhamedzhanov,richter14}: 
protons occupy only half of their crystallographic positions and, 
in the paraelectric phase, each of them tunnels back and forth 
between both sites of a double-well potential 
at a jump rate of the order of $10^{12}$s$^{-1}$ \cite{sugimoto}. 
Because the waiting time between jumps is larger 
by several orders of magnitude than the typical time of
x-ray resonant scattering ($\sim 10^{-15}$s), x-rays "see" the 
crystal as a series of snapshots, producing an effect similar to
thermal motion and static disorder \cite{dmitrienko2000}. 
Each transient proton configuration violates the crystal symmetry, 
but the space symmetry restores after averaging
over all possible proton configurations. 
A given proton configuration $\mathcal C$ induces a relaxation of the structure. 
The displacement $\bm u(\mathcal C)$ of the resonant atom from its high-symmetry site is accompanied by a correction to the resonant
scattering factor which is dominated by the E1E1 contribution
$\Delta f_{\alpha\beta}(\mathcal C)$. Similarly to the case of TMI scattering
\cite{dmitrienko2000}, it contributes to the resonant structure
factor with the partial contribution $f_{\alpha\beta\gamma}^{\mathcal C}$:
\begin{equation}
 f_{\alpha\beta\gamma}^{\mathcal C} = \Delta f_{\alpha\beta}(\mathcal C)u_\gamma(\mathcal C). 
 \label{fijkC}
\end{equation}
We can consider the global contribution of transient proton configurations $\bm f^{PC}$ 
as the coherent sum of the configurations $\mathcal C$ with probabilities $p(\mathcal C)$,
\begin{equation}
  \bm f^{PC} = \sum_{\mathcal C} p(\mathcal C) \bm f^{\mathcal C}
  \label{pc}
\end{equation}
Let us note that a given configuration $\mathcal C$ yields a nonzero 
contribution $\bm f^{\mathcal C}$ only if it induces a local
structure relaxation which displaces the resonant atom
from its high-symmetry site.

In the following, we will consider only three asymmetric proton configurations, 
which were proposed by Slater \cite{slater} and Takagi \cite{takagi2}:
the polar ($\mathcal P$), Slater ($\mathcal S$), and Takagi ($\mathcal T$) configurations (Figure \ref{fig.Hconfigs}). 
In the polar and Slater configurations, there are two protons near each $PO_4$
group, filling half of the four available sites. 
In the Takagi configurations, three protons are attached to one $PO_4$ group 
and only one proton is attached to a neighbour group.
In fact each of these three configurations may be decomposed into two similar 
and equiprobable configurations ($\mathcal P_1$ and $\mathcal P_2$, 
$\mathcal S_1$ and $\mathcal S_2$, $\mathcal T_1$ and $\mathcal T_2$)
whose sum entirely fills the crystallographic sites of the protons.
In the following, we consider their contribution by pairs, 
\textit{i.e.}, $\bm f^{\mathcal P} = \bm f^{\mathcal P_1} + \bm f^{\mathcal P_2}$, \textit{etc}.
Additionally, one should also consider the case of fully symmetrical configurations, 
when a PO$_4$ group is surrounded by either 0 or 4 protons. 
Due to their high symmetry, these configurations contribute to the dipole-quadrupole term only, 
and not to any extra term.

Altogether, the third-rank resonant atomic factor in the paraelectric phase of KDP can
thus be considered as the sum of five terms: 
\begin{equation}
 \bm f = \bm f^{dq} + \bm f^{TMI} + p(\mathcal P)\bm f^{\mathcal P} + p(\mathcal S)\bm f^{\mathcal S} + p(\mathcal T)\bm f^{\mathcal T}
 \label{f3}
\end{equation}

\subsection{Formalism of the structure factor in the paraelectic phase}

The formalism below applies only to the paraelectric phase of KDP, 
when the resonant atoms occupy the crystallographic sites with $\bar 4$ symmetry, 
\textit{i.e.}, the potassium (the experimental case presented here) or the phosphorus atoms.

Any third-rank tensor with $\bar 4$ point group symmetry 
admits six independent tensor components \cite{ITvolD}, 
but only three of them change sign under the glide-plane symmetry and contribute to the structure factor (\ref{F}) 
of the forbidden reflections of the type $hhl$ with $2h+l=4n+2$: 
$f_{xxz}$=-$f_{yyz}$, $f_{xzx}$=-$f_{yzy}$, and $f_{zxx}=-f_{zyy}$. 
It has been demonstrated \cite{mukhamedzhanov} that 
the structure factor (\ref{F}) of the dipole-quadrupole resonant scattering as well as those of 
the other resonant contributions considered in this paper 
can be written in the following matrix form:
\begin{equation}
  F \equiv F_{\epsilon\epsilon^\prime} = 
  \bm{\epsilon^\prime} .
  \left( \begin{array}{ccc}
  f_{xxz}H_z & 0 & f_s H_x+f_a L_x\\
  0 & -f_{xxz}H_z & f_s H_y+f_a L_y \\
  f_s H_x-f_a L_x & f_s H_y-f_a L_y & 0
 \end{array} \right)
 . \bm \epsilon
\label{sf}
\end{equation}
where $f_s=\frac 12 (f_{xzx}+f_{zxx})$, $f_a=\frac 12 (f_{xzx}-f_{zxx})$.

Except for the dipole-quadrupole term, all terms of Eq. \ref{f3} are of E1E1 resonance origin 
and are thus symmetric over permutation of the polarisation indices ($\alpha\beta$): 
thus only the dipole-quadrupole term may contribute to the antisymmetric part $f_a$.

It follows from Eq. \ref{sf} that different forbidden
reflections can have different energy spectra, since several
independent tensor components are involved in the structure
factor. In more detail, reflections $00l, l=4n+2$ are provided by
the $f_{xxz}$ component and the antisymmetric component $f_a$,
while reflections $hh0, h=2n+1$ are determined by the components
$f_s$ and $f_a$ (\textit{i.e.}, $f_{xzx}$ and $f_{zxx}$). All three
components contribute to the structure factor of $hhh, h=4n+2$
forbidden reflections.

\section{Experimental}
\label{exp}

\begin{figure*}
    \includegraphics[width=\columnwidth]{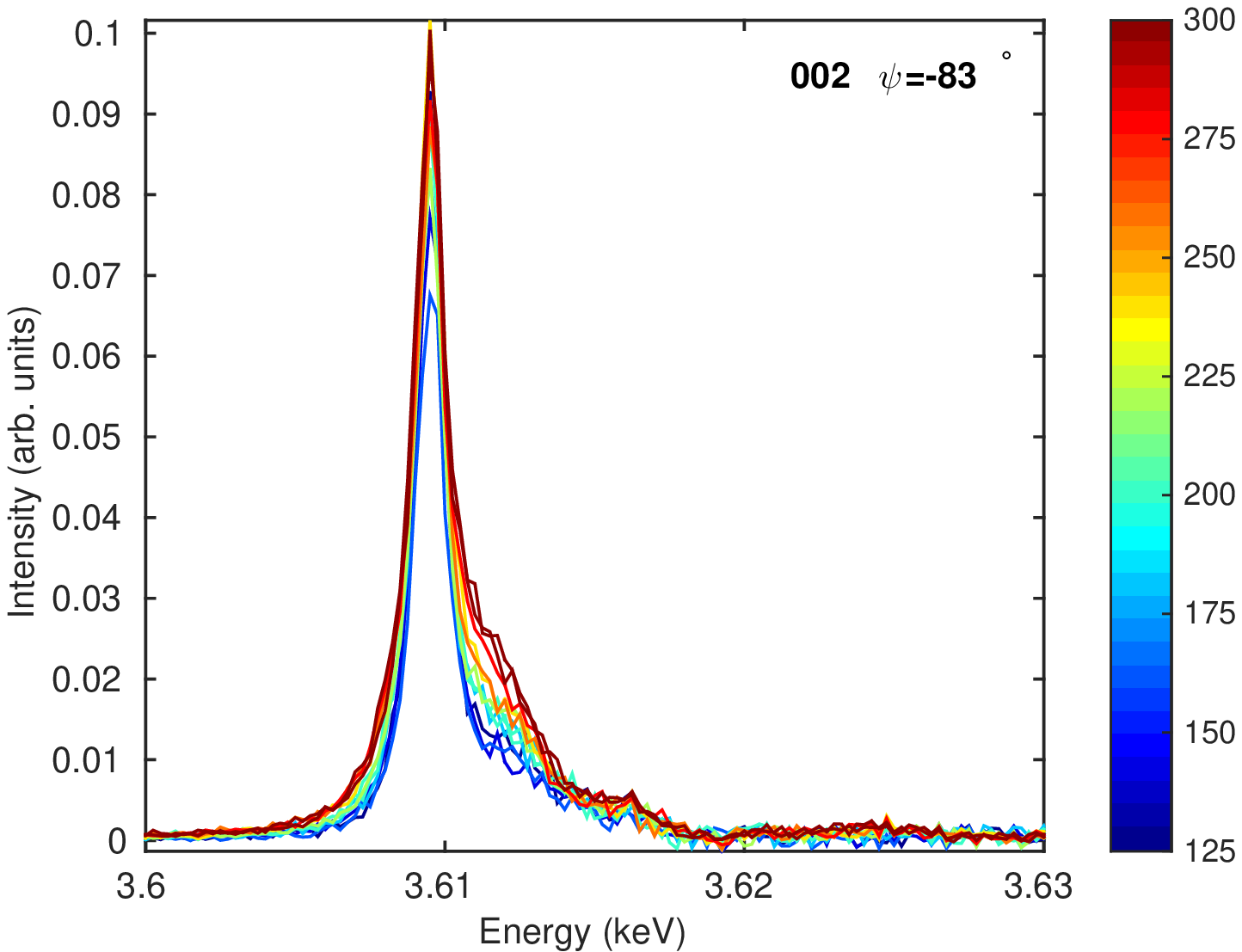}
    \includegraphics[width=\columnwidth]{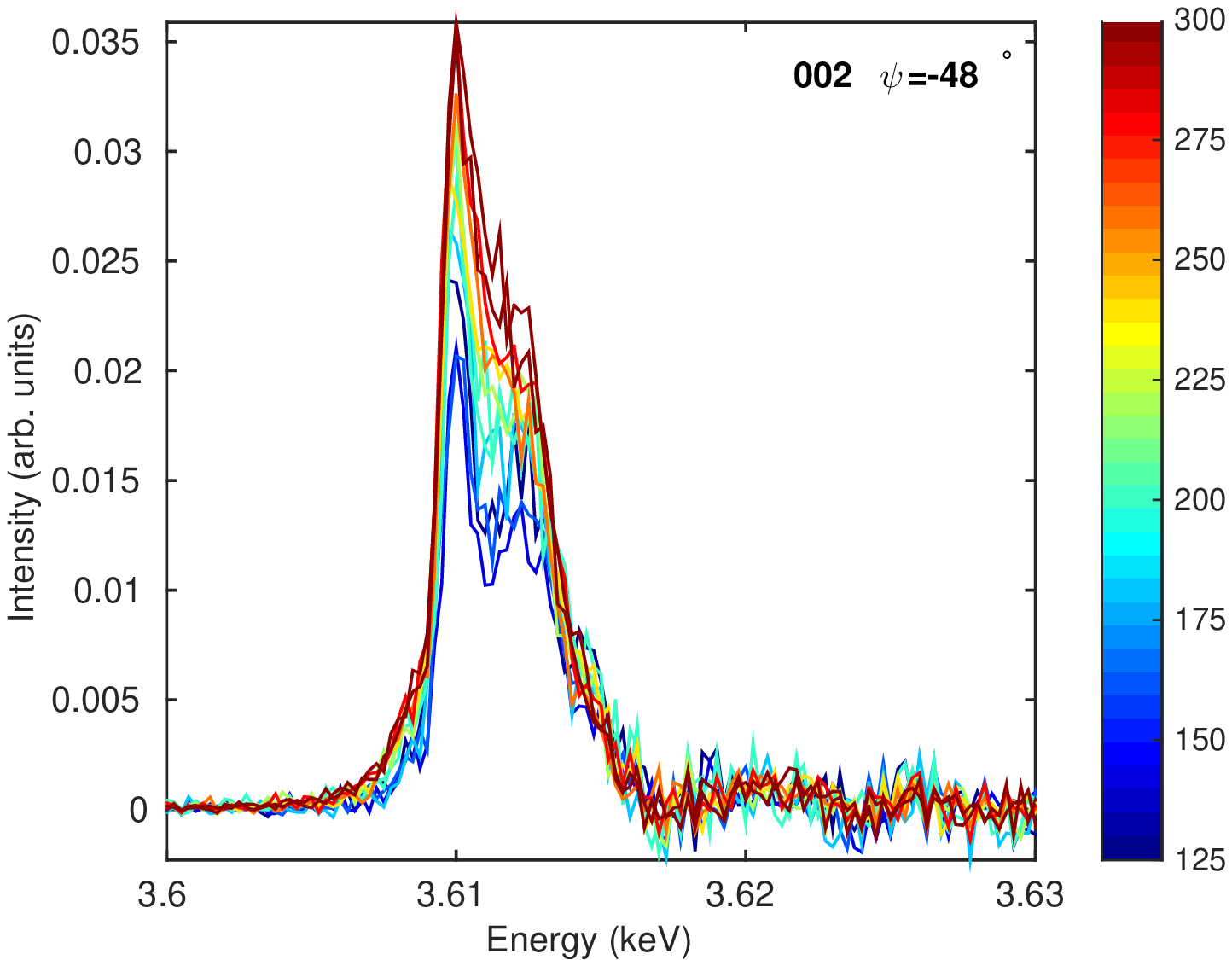}
    \includegraphics[width=\columnwidth]{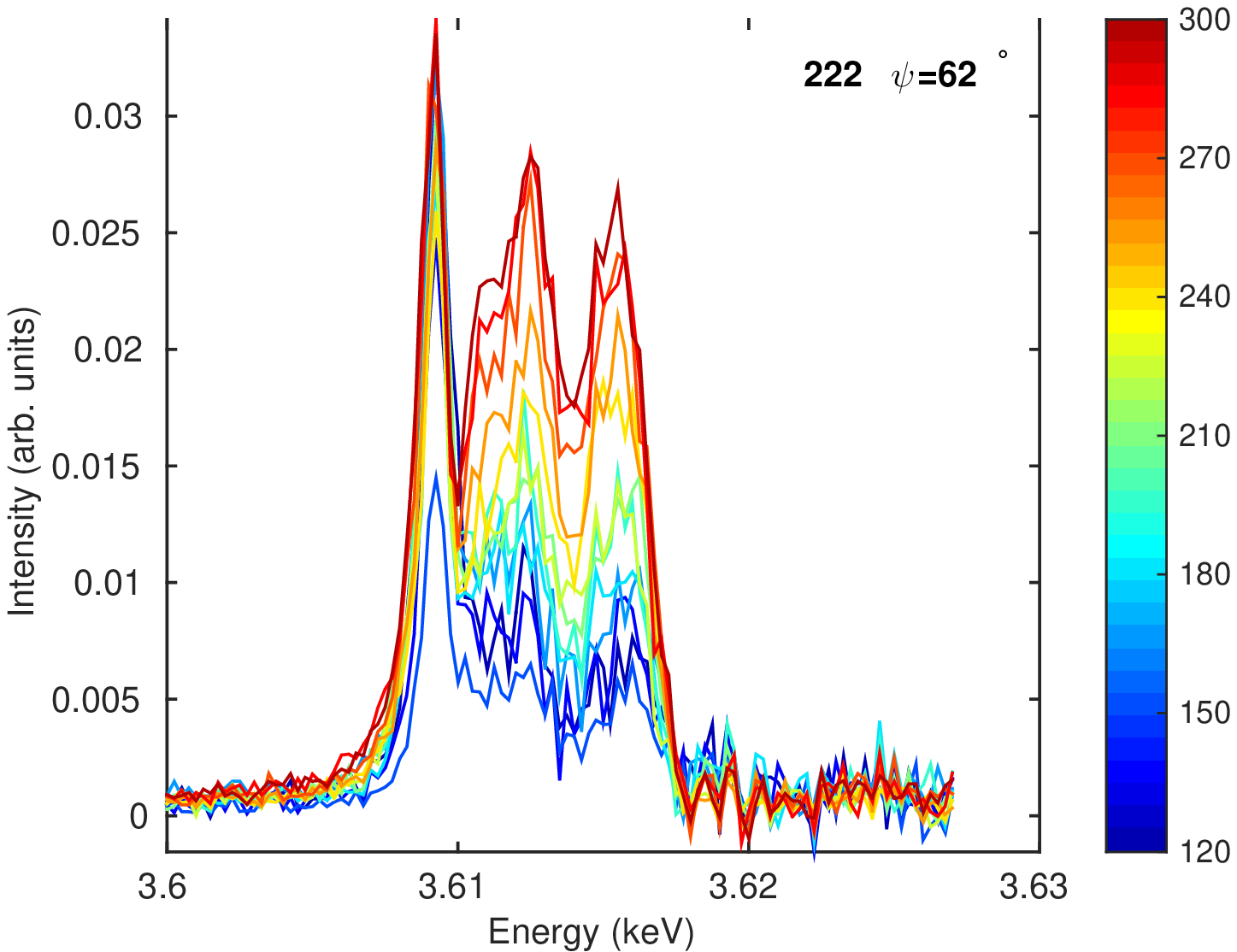}
    \includegraphics[width=\columnwidth]{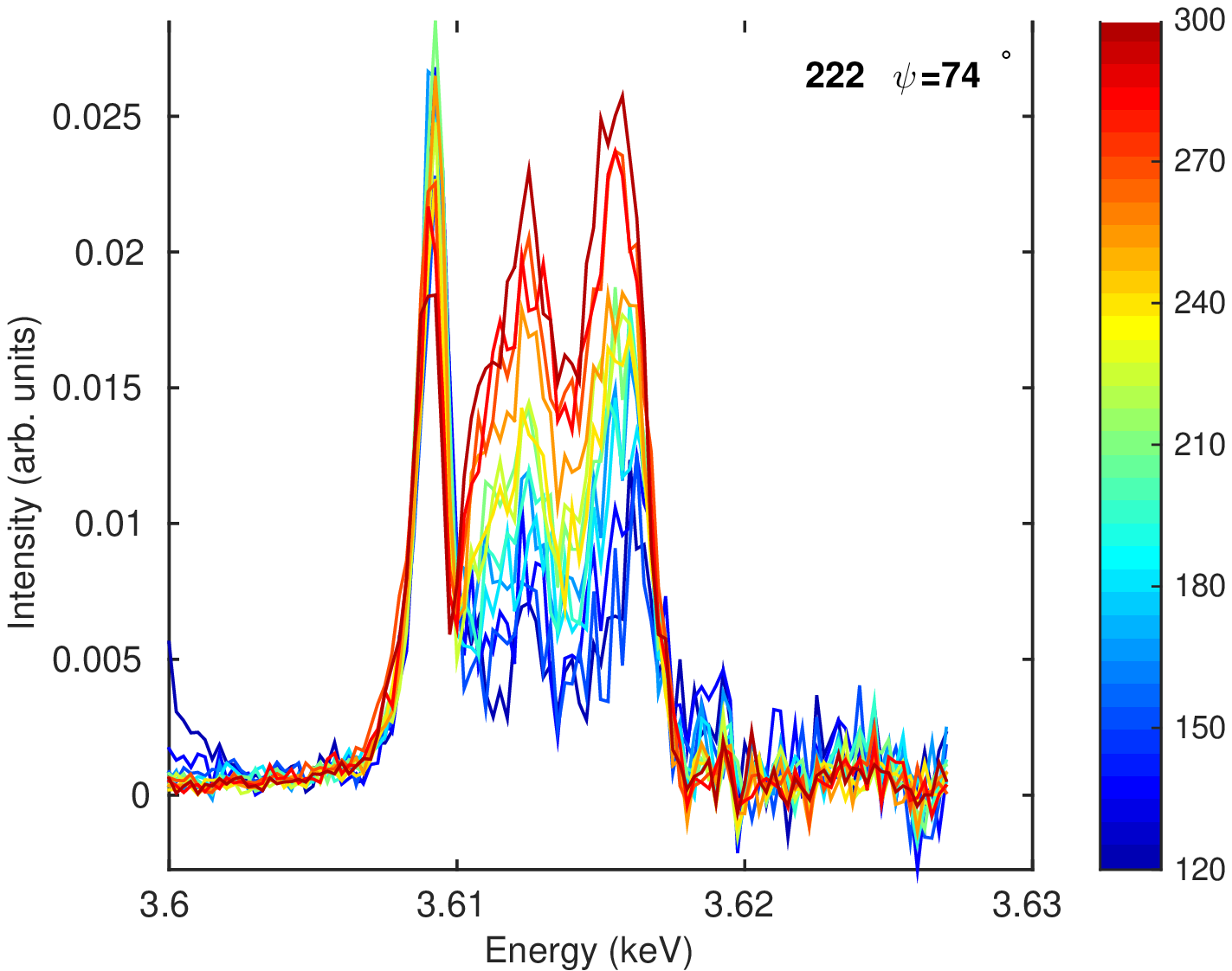}
    \caption{Temperature dependence of the spectra of the $002$ and $222$ forbidden reflections at two azimuths each. 
         Intensities were corrected for the fluorescent background, normalized by the incident beam intensity, and corrected for the varying ratio between integrated intensity and peak intensity of the rocking curve.
         Colour bars show the temperature scale.
         }
    \label{fig.TE}
\end{figure*}

A single crystal of KDP was grown and cut with surface normal 001 at the Institute of Crystallography (Moscow).
REXS was measured at beam line BM28 (XMaS) of the European Synchrotron Radiation
Facility, with preliminary measurements and fluorescence
measurements performed at beam line I16 of Diamond Light Source \cite{i16}. 
The sample was enclosed in a closed-cycle cryofurnace
and the temperature varied between 15 and 320 K. The $002$ and
$222$ forbidden reflections were measured at the potassium K edge
($\sim$3.608 keV). The measurements were performed in vertical
scattering geometry with the natural linear ($\sigma$)
polarisation of the incoming beam and without polarisation
analysis of the scattered beam. The azimuthal reference is the 100
axis and the azimuth $\psi$  is zero when the azimuthal reference is in the
scattering plane. KDP is known to suffer from radiation damage,
and so great care was taken to ensure the reproducibility of the
results presented here. Indeed, we observed radiation damage
during the preliminary measurements at beamline I16 when the incident beam was insufficiently attenuated.

Rocking curves were recorded at 3.6095 keV and showed a Lorentzian
shape with a varying width as a function of the temperature
\cite{beutier14}. Energy spectra were recorded in the same
temperature loop and were corrected for the fluorescence
background and for the varying ratio between integrated
intensity and peak intensity of the rocking curves. The corrected
spectra $I_{exp}(hkl,\psi)$ are shown in Fig. \ref{fig.TE}.

As reported in \cite{beutier14}, a spectacular change of spectrum and intensity can be seen across
the phase transition ({$T_c\approx123$ K}), due to the switching
on/off of the pure electric dipole (E1E1) component: the latter
vanishes in the tetragonal phase for symmetry reasons
\cite{beutier14,mukhamedzhanov}. In this paper we focus on the
tetragonal phase, whose energy spectra show interesting features
in their temperature dependence. Looking at the 002 reflection, we
see that 
(1) the spectrum changes with azimuth, meaning that more
than one independent component contributes to the structure factor,
in agreement with the symmetry analysis presented in Sec.
\ref{theory}; 
(2) the spectra at both azimuths change with temperature, revealing the contribution of more than one
scattering process, with different temperature dependences
(presumably one of them is independent of the temperature); 
and (3) the change of spectrum is stronger at $\psi=-83^\circ$ than at
$\psi=-48^\circ$. The case of the 222 reflection is less
spectacular but essentially shows the same features. We note that
both reflections have very different spectra, pointing at a very
different mix of the contributing amplitudes.

\section{Data analysis}

An analysis of the ferroelectric phase can be found in Appendix B. 
Here we deal only with the paraelectric phase.

The structure amplitude (\ref{sf}) of forbidden reflections involves three independent complex tensor components:
they interfere in the intensity and it is thus impossible to extract them directly from the four measurements (two reflections at two azimuths).
The analysis of the experimental spectra is therefore based on modelling with the FDMNES code \cite{bunau,fdmnes}.
The latter calculates resonant scattering amplitudes based on an input crystallographic configuration.
One should thus be able to evaluate several parameters of the crystallographic configuration, 
such as thermal motion and the relative probabilities of the various proton configurations, by trying to fit the experimental spectra.

In more details, we calculate the amplitudes $F_{\sigma\sigma}$ and $F_{\sigma\pi}$, 
which are the values of the structure factor (\ref{sf})
for incident polarisation $\sigma$ and scattered polarisation $\sigma$ and $\pi$ respectively.
The calculations are performed with the multiple scattering method of FDMNES \cite{joly}, 
using the convolution parameters obtained from the fits of the absorption spectra (see Appendix A).

\subsection{Self-absorption correction}

In the kinematical theory of diffraction, 
the integrated intensity $I$ measured in Bragg geometry from a thick sample, 
with incident polarisation $\sigma$ and no polarisation analysis of the scattered beam, 
is proportional to
\begin{equation}
  I = \left[ \frac{\mid F_{\sigma\sigma}\mid^2}
  {\mu_\sigma+g\mu^\prime_\sigma} + \frac{\mid F_{\sigma\pi}\mid^2}
  {\mu_\sigma+g\mu^\prime_\pi} \right]e^{-2M} \label{int}
\end{equation}
where $\mu$ and $\mu^\prime$ are the polarisation-dependent absorption coefficients of the incident and outgoing beams respectively, 
$g=\frac{\sin\eta}{\sin\eta^\prime}$ is a geometrical factor given by the incident angle $\eta$ and exit angle $\eta^\prime$ with respect to the sample surface,
and $e^{-2M}$ is the Debye-Waller factor.
This expression takes into account the anisotropic absorption of the material, 
provided it is small enough so that the polarisation of x-rays is not modified along the propagation.
In the following we will use a single absorption coefficient $\tilde\mu$ 
instead of three distinct ones, leading to the simplified expression:
\begin{equation}
  I \approx \frac {\mid F_{\sigma\sigma}\mid^2+\mid F_{\sigma\pi}\mid^2} {\tilde\mu(1+g)}  e^{-2M}
  \label{intensity}
\end{equation}
The anisotropic character of the material is reflected by the
choice of $\tilde\mu$, which is chosen differently for different
reflections and different azimuths. For instance, for the $002$
reflection, which is parallel to the tetragonal axis,
$\tilde\mu=\mu_\perp$, where $\mu_\perp$ is the absorption
coefficient for a beam with polarisation perpendicular to the
tetragonal axis, is a good approximation (for all azimuths), since
in this particular case $\mu_\sigma=\mu^\prime_\sigma=\mu_\perp$.
On the other hand, for the 222 reflection at the azimuths reported
here, it turns out that $\tilde\mu=\mu_{iso}$, where $\mu_{iso}$ is the isotropic part 
of the linear absorption coefficient, is a reasonable approximation.

The experimental spectra $I_{exp}(E)$ were fitted against Eq. (\ref{intensity}), 
where $e^{-2M}$ is taken from the literature \cite{usquare} 
and the spectra $F_{\sigma\sigma}(E)$ and $F_{\sigma\pi}(E)$ 
are calculated according to a procedure detailed below.

\subsection{Model of the resonant scattering amplitudes}
\label{calc}

According to the model described in Sec. \ref{theory}, 
the resonant scattering factor is the sum of the
dipole-quadrupole, TMI and various PC contributions [Eq. (\ref{f3})].
Methods to calculate the various contributions to the atomic scattering factor 
were developed in previous works \cite{ovchinnikova10,richter14}. 
It is, however, easier to work directly on the structure amplitudes, 
which are a linear function of the tensor components of the atomic scattering factor (equation \ref{sf}).
We can write:
\begin{equation}
 F_{\epsilon\epsilon^\prime} (hkl,\psi,E,T) = 
  \sum_X a^X (T) F_{\epsilon\epsilon^\prime}^X (hkl,\psi,E)
 \label{SF}
\end{equation}
where the $F_{\epsilon\epsilon^\prime}^{X}$ ($X\in\{dq,TMI,\mathcal{P,S,T}\}$, 
$(\epsilon,\epsilon^\prime)\in\{\sigma,\pi\}$) 
are the structure factors of the various contributions 
projected onto the polarisation states according to equation \ref{sf}. 
We will justify below that the temperature dependence can be fully accounted for in the $a^X$ coefficients, which in turn are independent of the other parameters.
In the case of the contributions of the proton configurations, the $a^{X}$ coefficients 
are proportional to the probability of the corresponding configurations.

The dipole-quadrupole (E1E2) contribution can be calculated directly from the average crystal structure with FDMNES, 
while preliminary modelling is required for the other terms.
\textit{Ab initio} calculations showed that its spectrum in the wurtzites 
is essentially temperature independent \cite{oreshko12}.
In the case of KDP, we find the same result, using the temperature-dependent structure proposed in Ref. \onlinecite{nelmes1982}.
Nevertheless, we allow for a global temperature-dependent scaling factor $a^{dq}$, 
which accounts for a small dependence of the E1E2 term on atomic positions.

\subsection{Model of TMI scattering}

The TMI contribution to the resonant structure factor was calculated 
with the same method as that developed for Ge and the wurtzites \cite{ovchinnikova05,ovchinnikova10}, 
which has been validated by \textit{ab initio} calculations \cite{oreshko12}.
We simulated a $2\times 2\times 2$ supercell
in which all atoms were randomly displaced from their average positions. 
The displacement amplitudes were chosen according to the data given in Ref. 
\onlinecite{nelmes1982} for {125 K}. 
This model neglects the correlations between the displacements of the various atoms.
Therefore it yields similar results to a model taking into account the displacements 
of the resonant atoms alone.
In this approximation, the TMI contribution to the atomic factor 
depends linearly on the displacements of the resonant atom, 
which are supposed to be isotropic. 
Moreover the calculations of the TMI spectrum show that 
its lineshape does not change with temperature, 
so that the temperature plays only as a global scalar on the spectrum.
This point has been demonstrated with \textit{ab initio} calculations 
in the case of ZnO and GaN \cite{oreshko12}.
We can thus consider the TMI contribution as the product of a temperature-independent 
spectrum $F_{\epsilon\epsilon^\prime}^{TMI}(E)$ 
and a temperature-dependent scalar coefficient $a^{TMI}(T)$.

\subsection{Model for the contribution of the proton configurations}

A method to calculate the PC contributions has been reported \cite{richter14}: 
it consists of calculating the various tensor components $f_{\alpha\beta\gamma}^{\mathcal C}$ 
for each configuration according to Eq. (\ref{fijkC}) after relaxing 
the structure in the chosen configuration.
After the calculation of the $f_{\alpha\beta\gamma}^{\mathcal C}$,
phenomenological expressions describing the azimuthal dependence of the reflections
are derived and used to fit the experimental data.
This approach gives satisfactory fits to the experimental data. 

In order to improve the spectra description we present here another 
method to simulate the energy spectra, 
which we believe is more reliable because all
calculations are made in the same FDMNES calculation
and there is no necessity to use phenomenological expressions.
Instead of modelling the Cartesian components $f^\mathcal{C}_{\alpha\beta\gamma}$, 
we directly calculate the structure amplitudes $F_{\epsilon\epsilon^\prime}$ 
of the forbidden reflections as a sum of several contributions. 
Similarly to the method applied in Ref. \onlinecite{ovchinnikova10} 
we suppose that each contribution to the scattering amplitude 
may be considered as a temperature-independent spectrum 
$F^X_{\epsilon\epsilon^\prime}(E)$ ($X\in\{\mathcal{P,S,T}\}$) 
multiplied by a temperature-dependent coefficient $a^{X}(T)$. 
To calculate the PC contributions, 
we have constructed supercells, where all protons 
occupy the same configuration chosen among the $\mathcal P$, $\mathcal S$, and $\mathcal T$ configurations. 
Here we consider the proton configurations as independent static defects. 
The {\it ab initio} code VASP \cite{kresse96,kresse99} 
was used to compute the relaxation of the structure
for the chosen proton configuration and obtain new atomic coordinates. 
The resulting coordinates are slightly different inside each
pair of configurations. 
Of particular interest is the displacement $\bm u(\mathcal C)$ of the resonant atom.
These calculations were made for each temperature, 
taking into account the change of lattice parameters
and of the spacing $\delta$ between the two proton sites of the H-bond, 
in correspondence with the data given in Ref. \onlinecite{nelmes1982}. 
The variation of $\bm u(\mathcal C)$ obtained in the {125-300 K} range does not exceed 5$\%$ 
for the three types of proton configurations considered here 
and provides similarly weak variations of $\Delta f_{\alpha\beta}(\mathcal C)$.
The variation is essentially a global scaler of the spectrum with a linear dependence in temperature. 
It supports our model, in which each contribution $F^X_{\epsilon\epsilon^\prime}(E)$ 
is assumed to be independent of the temperature, 
provided the global dependence is included in the coefficient $a^X(T)$.
Then the spectra $F^X_{\epsilon\epsilon^\prime}(E)$ 
corresponding to each configuration are calculated with FDMNES, using the relaxed structures.
The coefficients $a^{X}(T)$ are obtained by fitting 
the experimental data at each recorded temperature.

\begin{figure*}
    \includegraphics[width=\columnwidth]{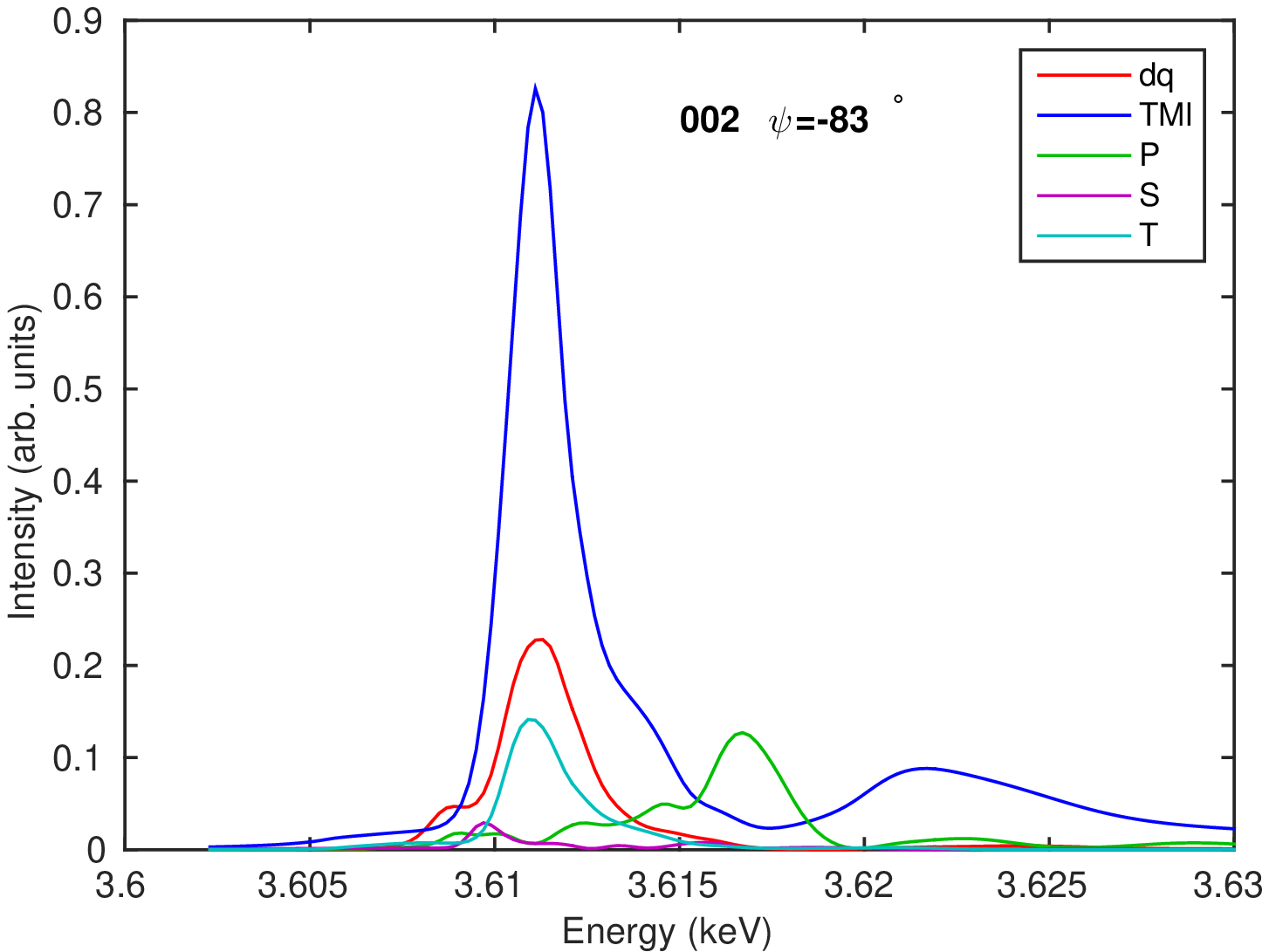}
    \includegraphics[width=\columnwidth]{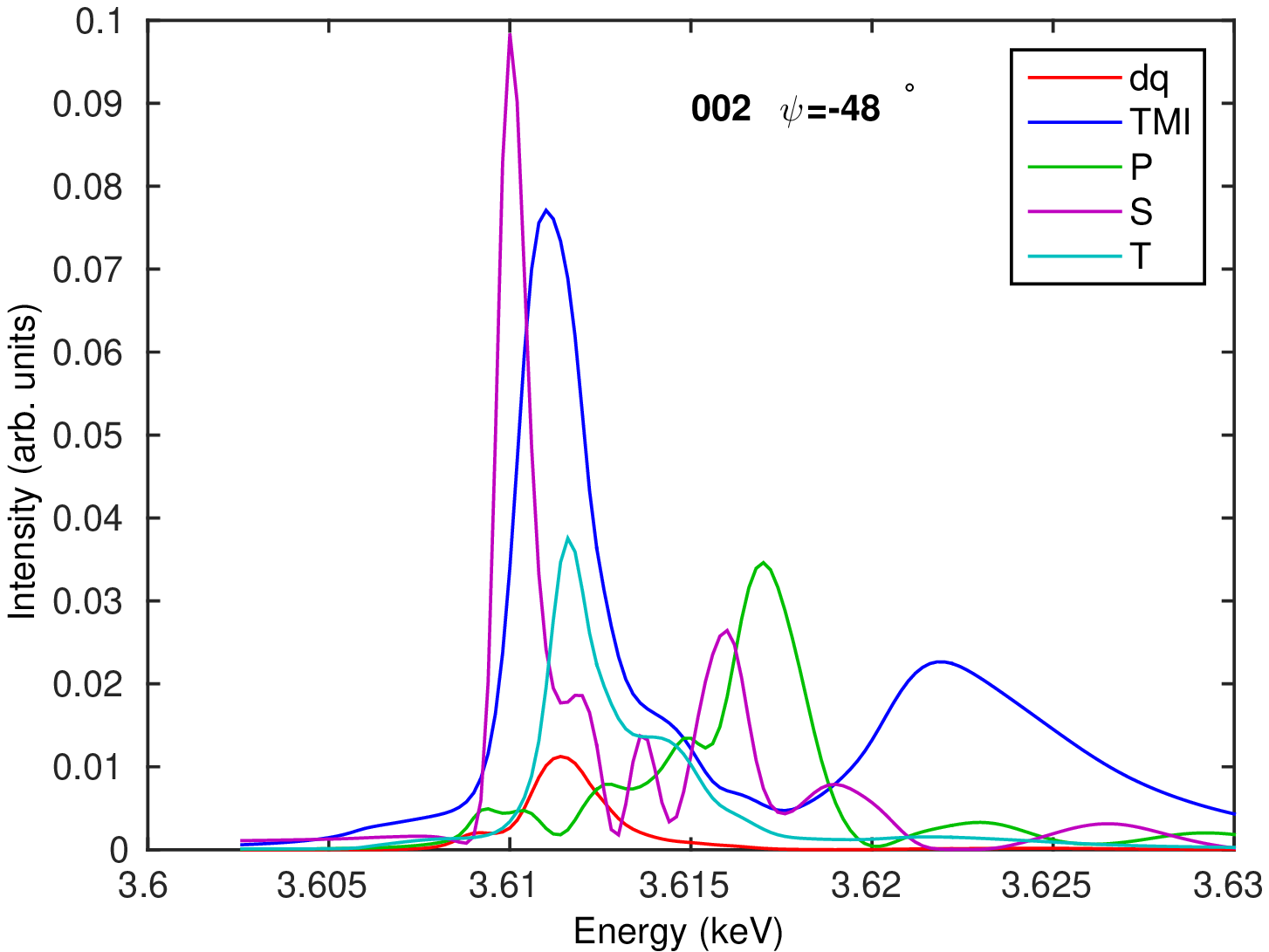}
    \includegraphics[width=\columnwidth]{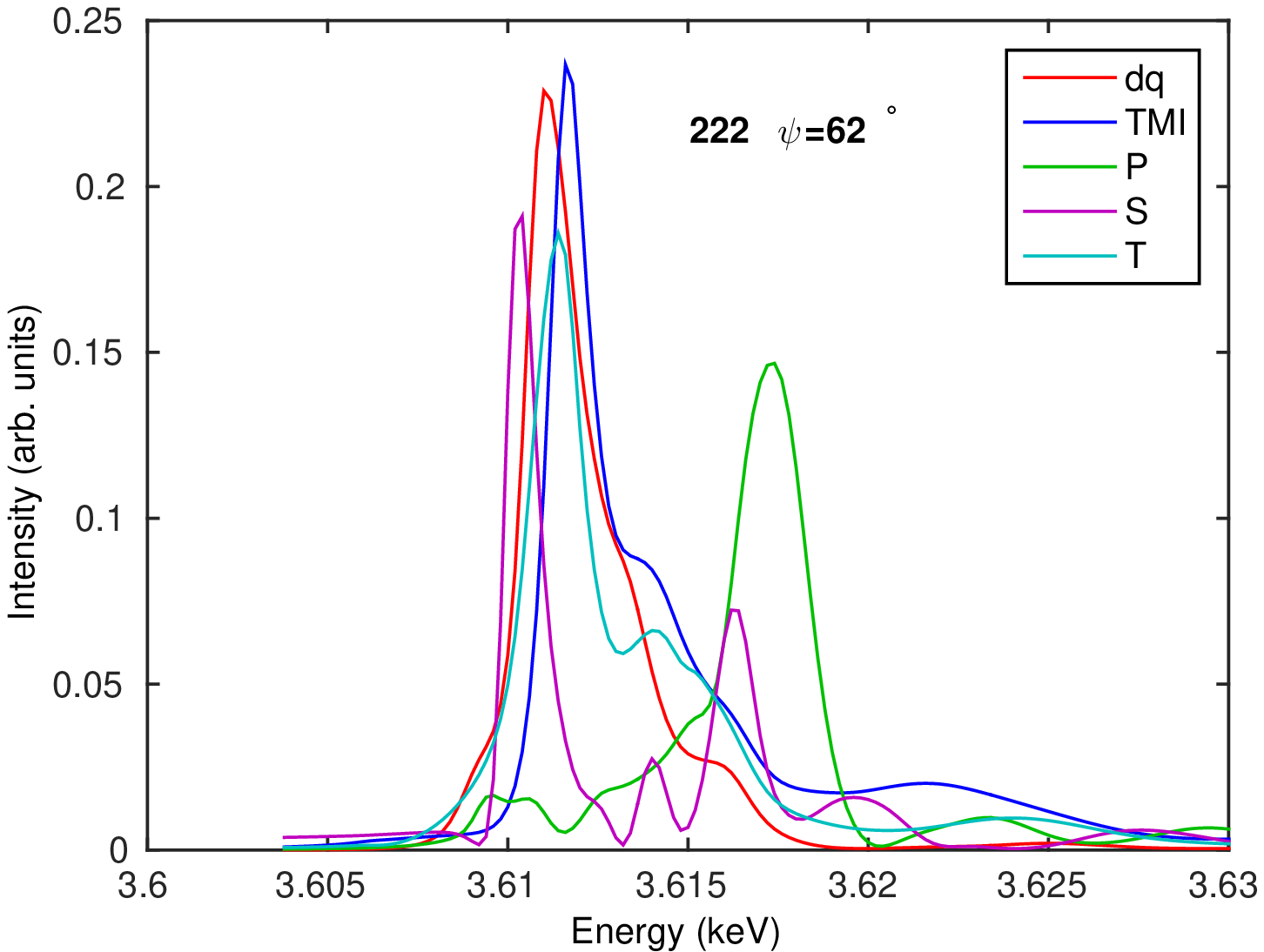}
    \includegraphics[width=\columnwidth]{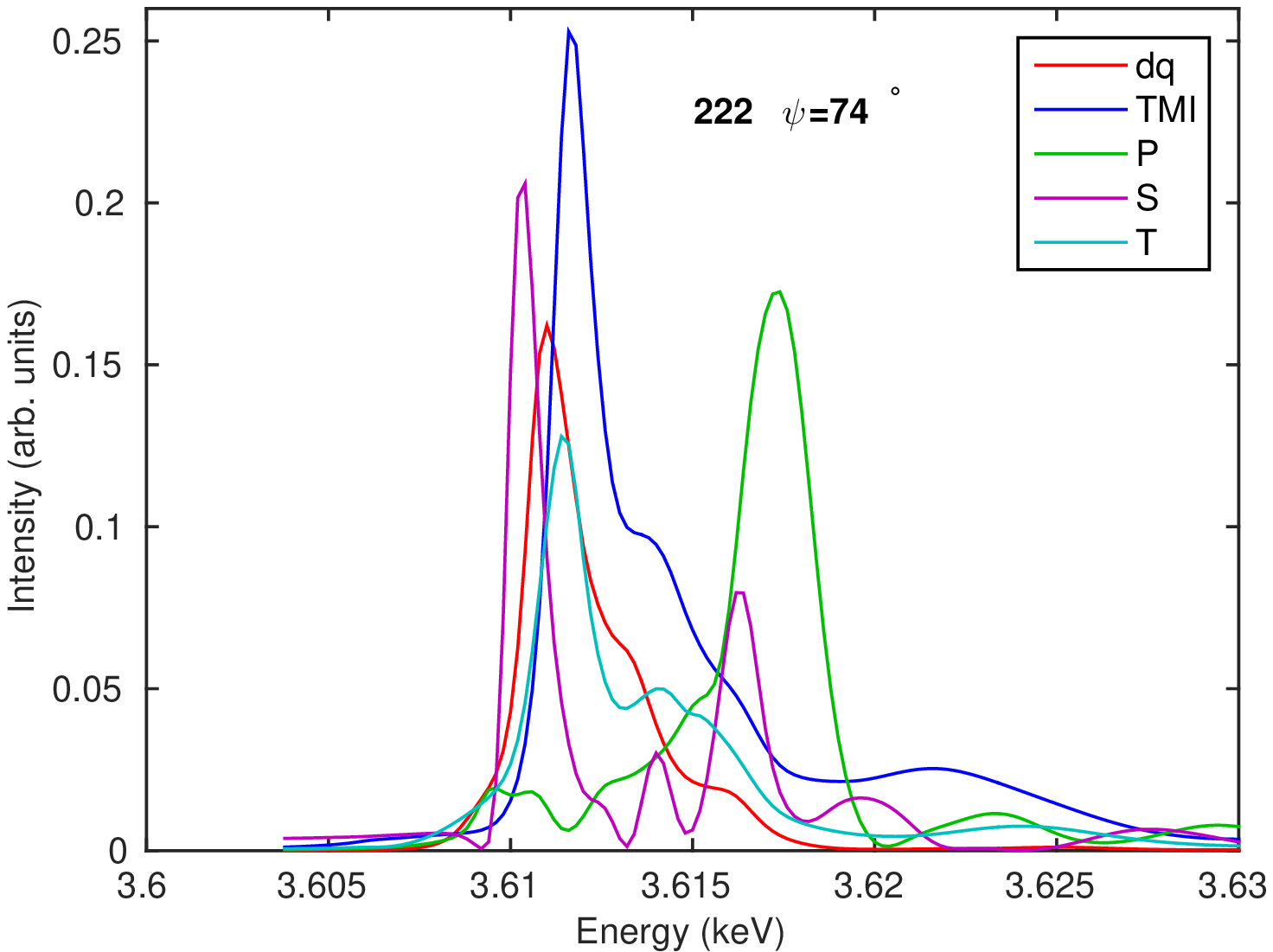}
      \caption{Contributions of the various scattering processes to the intensity of the forbidden reflections $002$ and $222$, for the measured azimuthal angles, at 300 K.
      See text for details.}
    \label{contributions}
\end{figure*}

\subsection{Spectral contributions and fits}

For each temperature, we have four experimental energy spectra: 
two reflections at two different azimuths each.
The four scalar parameters of the model can thus be reliably 
determined by fitting the experimental spectra.
The task is simplified because all contributions possess different energy
dependences: it is more or less obvious which contribution is
responsible for different parts of the energy spectra. 
This is shown in Fig. \ref{contributions}, which presents the contributions
$I^{X}(E)=\left(|F^X_{\sigma\pi}(E)|^2+|F^X_{\sigma\sigma}(E)|^2\right)/\tilde\mu(E)$
for the $002$ and $222$ reflections each for two azimuthal angles at 300 K. 
We see that the dipole-quadrupole, TMI, and Takagi configurations 
contribute in the structure factor mainly in the lower part of the energy spectra, 
while the higher-energy side is mainly provided by the polar and Slater configurations. 
Moreover, the dipole-quadrupole and TMI contributions predominate in the $002$ reflection, 
while the PC contributions become more important in the $222$ reflection.

Nevertheless, the intensity spectra of the forbidden reflections 
are not simple sums of these partial intensity spectra, 
but are determined by the interference between the complex amplitudes.
By fitting the experimental spectra against Eqs. (\ref{intensity}) and (\ref{SF}), we obtained the coefficients $a^X$. 

In Ref. \onlinecite{richter14}, only the Slater configurations were evidenced:
the polar configurations were found to contribute to the experimental spectra 
and the Takagi configurations were neglected due to their higher energy. 
In the case of KDP, it turns out that all three types of configurations are needed 
to explain the experimental spectra, which are much more complex than those of RDP. 
However, the line shape of the Takagi configurations cannot easily be disentangled from that of the TMI and polar configurations.
While they provide a sensible improvement of the fits, their coefficient $a^\mathcal{T}$ lacks reliability.

\subsection{Results}

\begin{figure*}
    \includegraphics[width=\columnwidth]{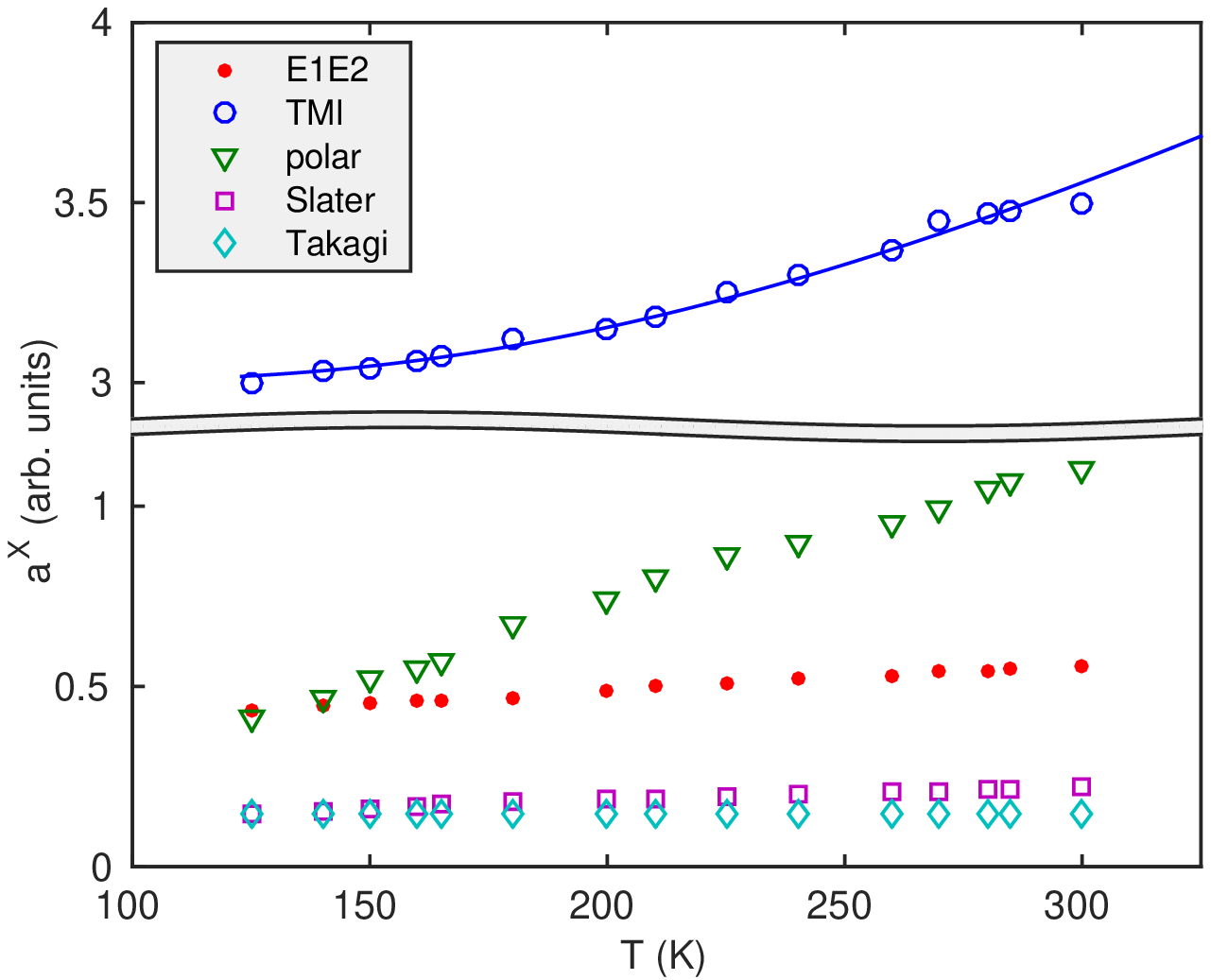}
    \includegraphics[width=\columnwidth]{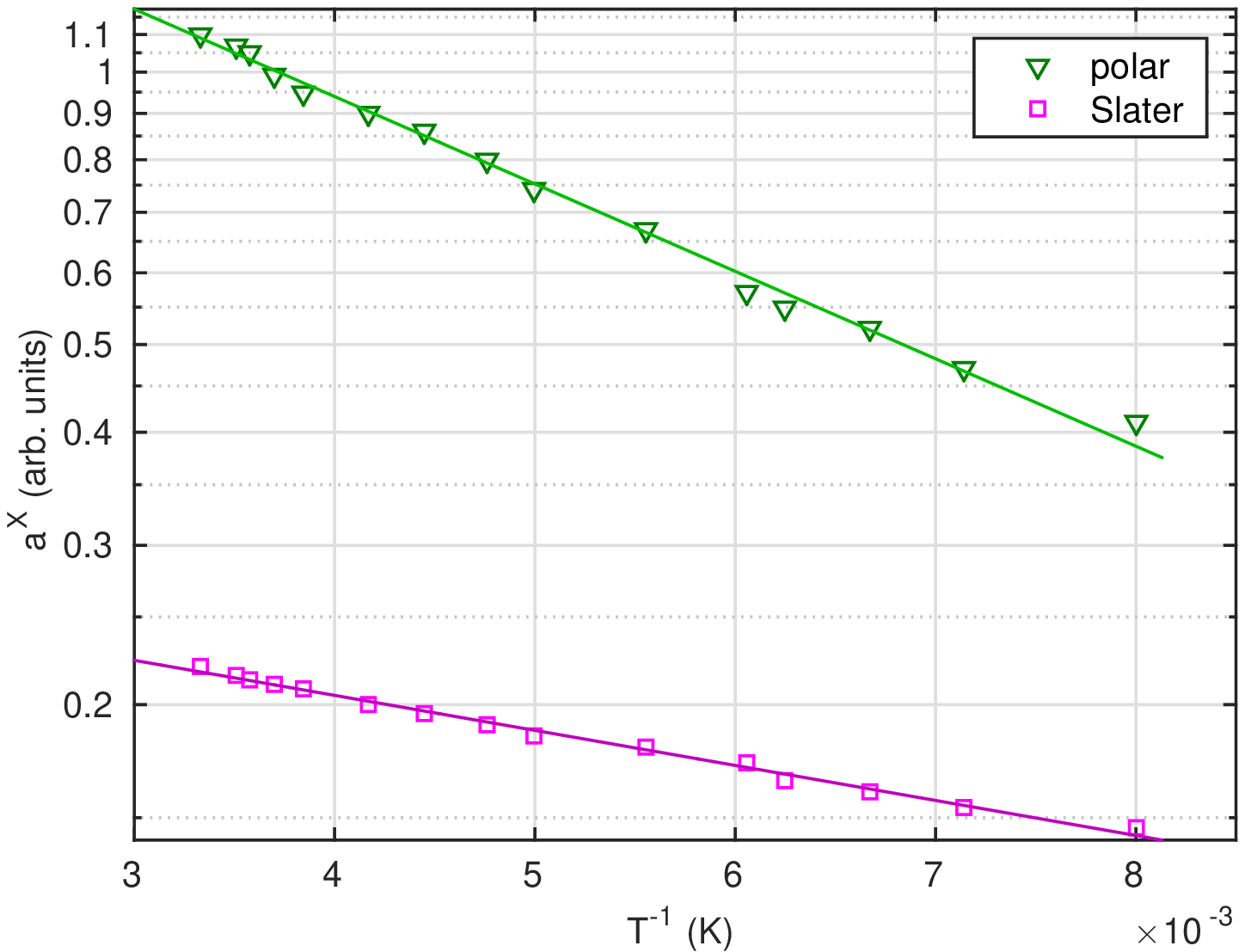}
         \caption{Left: Temperature dependence of the various contributions to forbidden reflections. 
         Right: Arrhenius plots of the polar and Slater contributions.
         (Open circles: experimental data obtained from the spectra fitting; solid line: approximating functions (see text for details).}
    \label{fitting}
\end{figure*}

\begin{figure*}
    \includegraphics[width=\columnwidth]{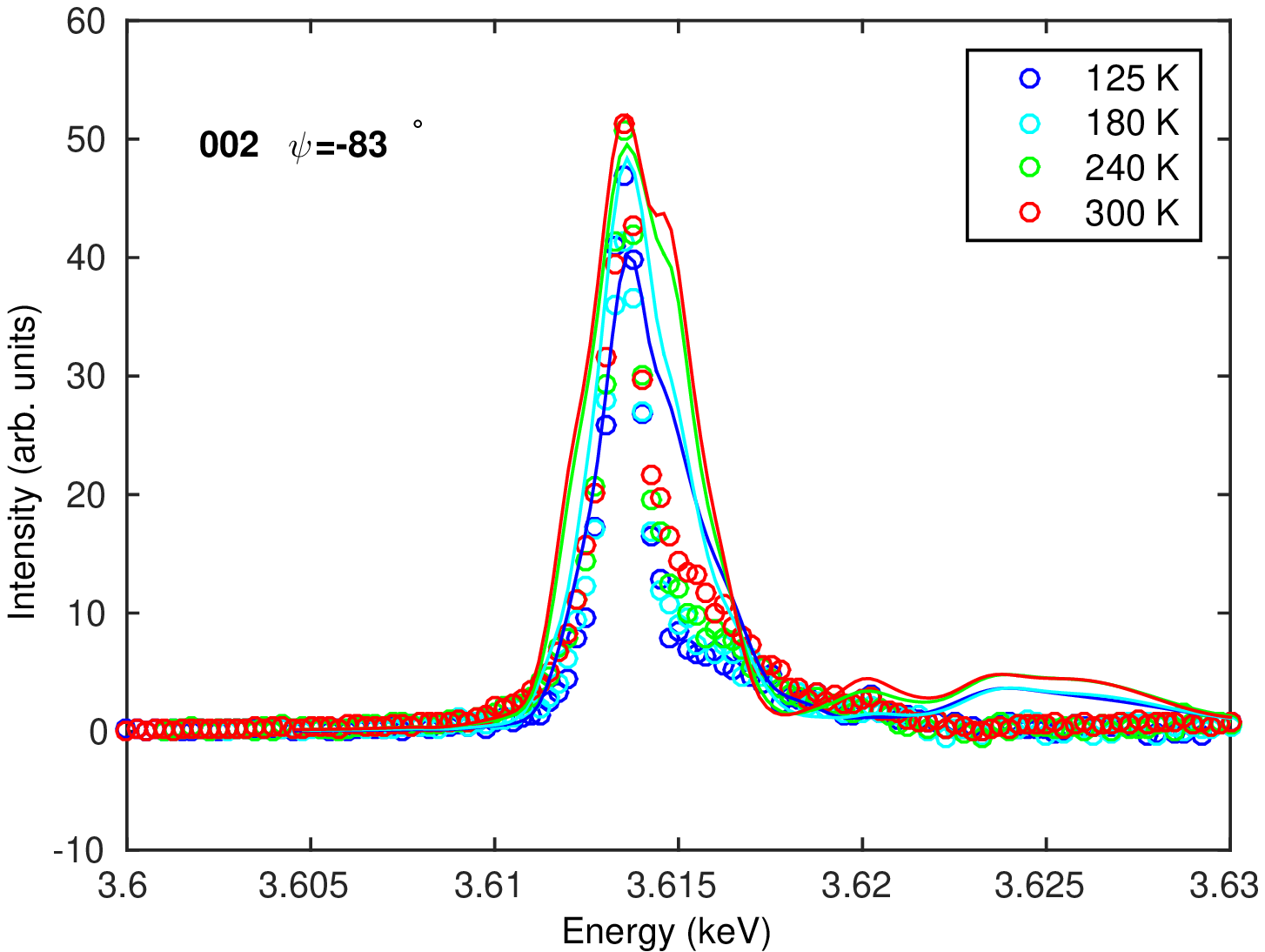}
    \includegraphics[width=\columnwidth]{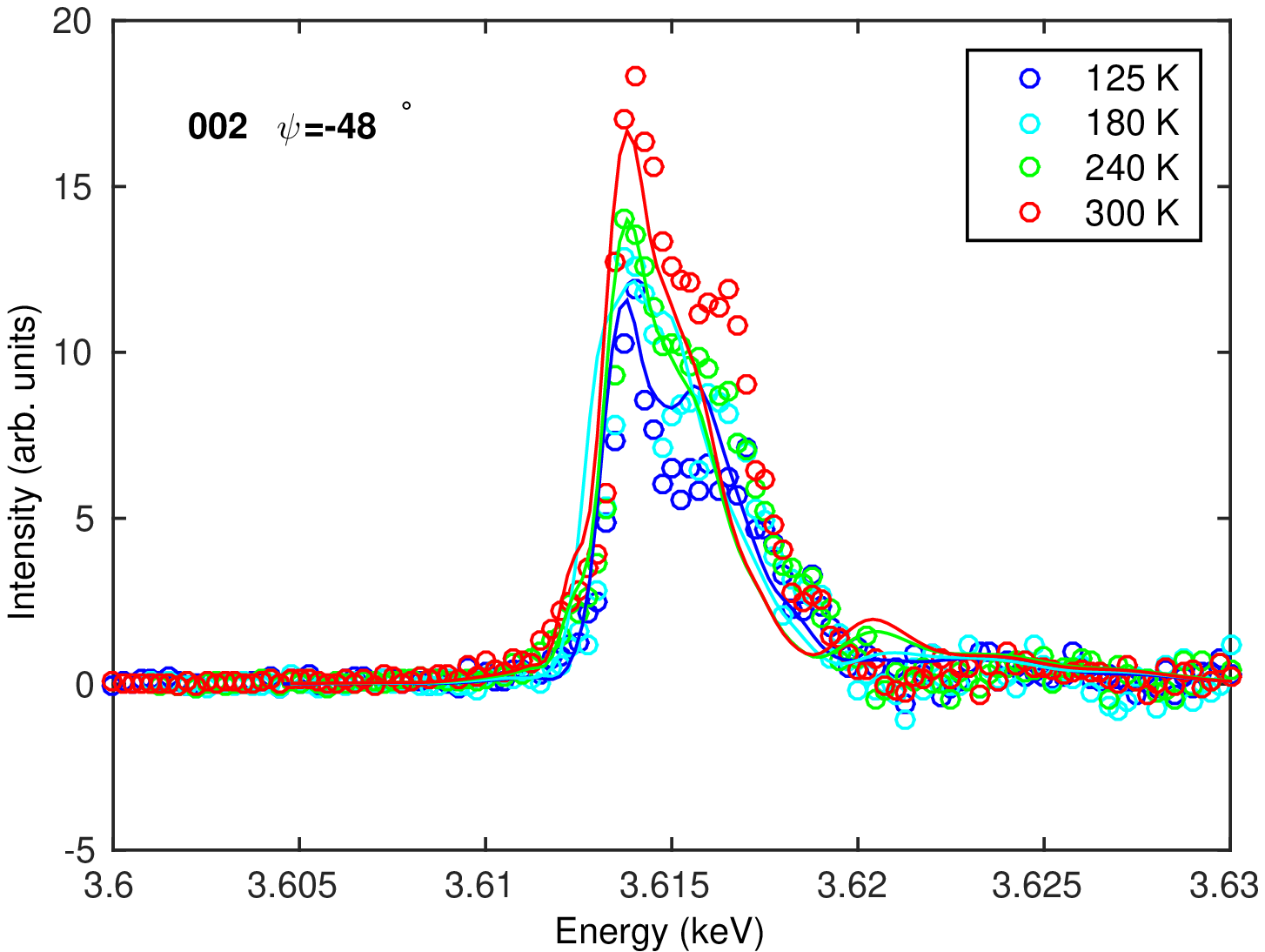}
    \includegraphics[width=\columnwidth]{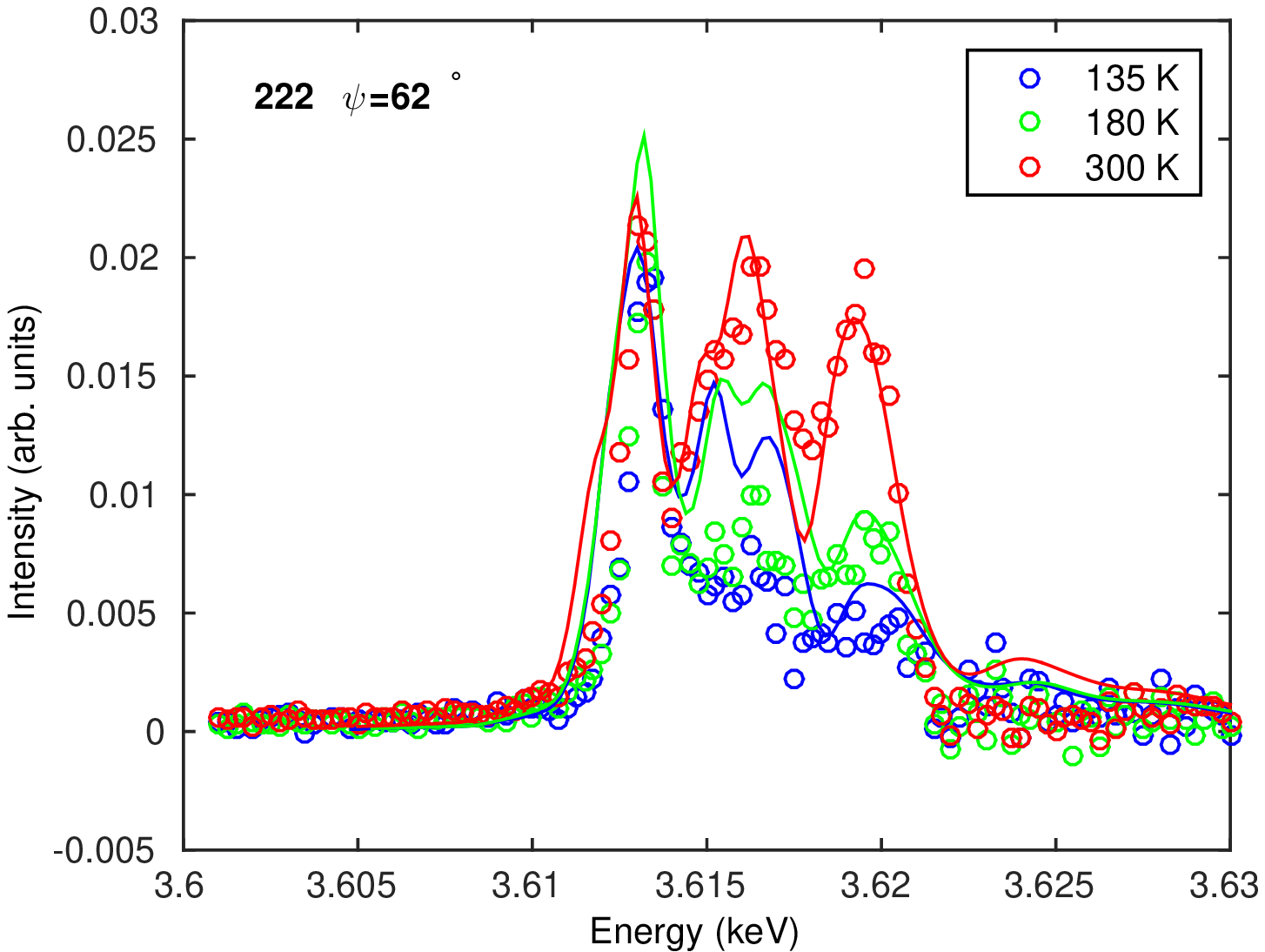}
    \includegraphics[width=\columnwidth]{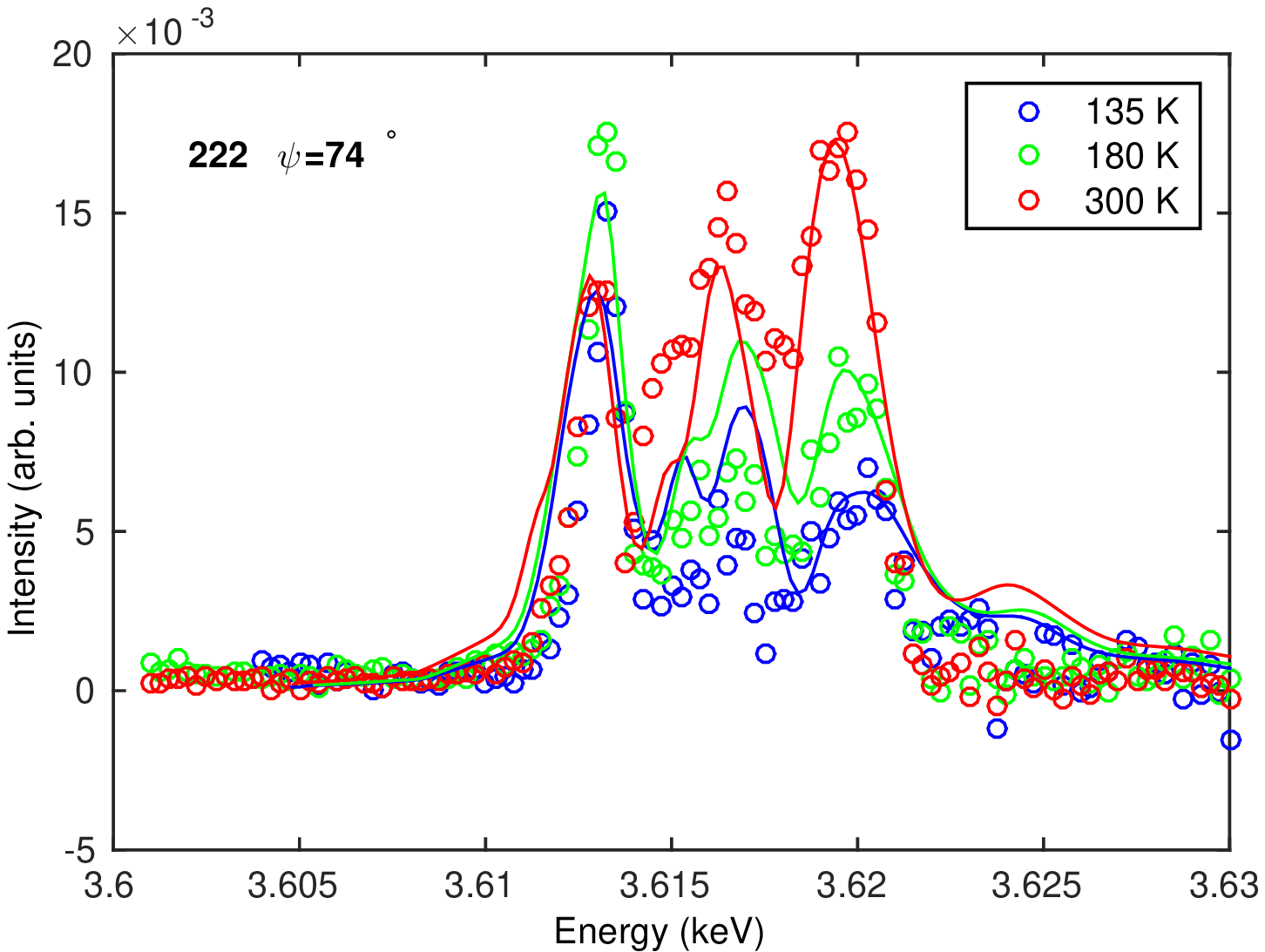}
      \caption{Energy spectra of the $002$ and $222$ reflections at the two measured azimuths,
      at various temperatures. Open circles: experimental data; solid lines: calculations.}
    \label{fits}
\end{figure*}

The fits were performed at each measured temperature,
providing the temperature dependence of the coefficients $a^X(T)$ (Fig. \ref{fitting}).

We see that the spectra are dominated by the TMI amplitude at all temperatures. 
However, its thermal growth is rather weak compared to previously reported cases \cite{kirfel2002,collins03,beutierGaN}.
Nelmes \textit{et al.} found a doubling of the thermal parameters of the potassium atoms between the phase transition and room temperature \cite{nelmes1982}, 
while the TMI growth that we observed in the same temperature range is much weaker. 
This can be explained by the fact that only part of the optical vibration modes contributes to the TMI effect.
We fit the TMI growth with the usual phonon statistics model, assuming a single optical mode \cite{kokubun2001}: 
\begin{equation}
 a^{TMI}(T) = a^{TMI}_0 \coth\left(\frac{E^{TMI}}{k_B T}\right)
\end{equation}
We find $E^{TMI}=32$ meV $\equiv 258$ cm$^{-1}$. 
This value does not correspond to any reported vibration mode of KDP, but there are several modes between 150 and 500 cm$^{-1}$ \cite{she}, 
such that our single mode model is a too crude approximation in this case.

The dipole-quadrupole term also slightly grows with temperature, like in Ge \cite{oreshko11}. 

It is more interesting to explain the temperature dependence of the PC contributions. 
In accordance with Eq. (\ref{pc}) we believe that $a^{PC}$ scales with the number of defects, which presumably follows the Arrhenius law: 
\begin{equation}
 \ln(a^X(T))=-\frac{E^X}{k_B T}+\ln(a^X_0)
\label{Arrhenius}
\end{equation}
where $E^X$ is the activation energy and $a^X_0$ is a constant. This constant contains various scaling factors related to the scattering measurements.
Figure \ref{fitting} shows that the polar and Slater configurations indeed follow this law.
From the linear fits we extract their activation energies:  $E^\mathcal{P}=18.6\pm  0.5$ meV and $E^\mathcal{S}=7.3\pm 0.2$ meV.
The determination of the $a^\mathcal{T}$ is not reliable enough to allow further analysis of the Takagi configurations.

In KDP, the polar configuration is usually considered as the ground state because it corresponds to the low temperature ferroelectric phase. 
The Slater and Takagi configurations are considered as first and second excited states respectively \cite{fairall,lasave2005}.
The situation is reversed, for instance, in ammonium dihydrogen phosphate (ADP), in which the antiferroelectric order is stabilized by the Slater configurations \cite{nagamiya,hewat,lasave2007}. 
Our results are therefore inconsistent with the usual model. 
It is clear that this result must be taken carefully since the best fits reproduce only roughly the measured spectra (Fig. \ref{fits}) and that some spectroscopic features are inaccurate. 
The quality of the fits is nevertheless standard for REXS spectra.
The discrepancy is likely to originate from the evaluation of the contributing spectra, for two main reasons. 
First, the structural model of each proton configuration is rather simple: 
each configuration is simulated as a crystal of identical configurations, thus ignoring the possible interplay between different neighbour configurations. 
Moreover, correlations are also ignored in the simulation of the thermal effects.
Second, the spectroscopic calculations of forbidden reflections are never very accurate, even in simple cases, 
because they involve complicated quantum calculations, such as the convolution with the width of the excited state, which is unknown. 
For these two reasons, it is not impossible that the contribution of the polar configurations is not well evaluated. 
Nevertheless, we found, maybe fortuitously, a value of the activation energy of Slater configurations (7.3 meV) that is in fair agreement with the values reported in the literature.
The activation energy of Slater configurations with respect to the polar configurations has been evaluated to 5.2 meV by Fairall and Reese \cite{fairall}, 
based on a phenomenological model and experimental polarisation curves.
Recent \textit{ab initio} calculations yielded values of 16.9 meV for uncorrelated Slater defects and 5.0 meV for Slater defects correlated in chains \cite{lasave2005}.
The better agreement of the latter with the value of Fairall and Reese suggests the occurrence of correlated Slater clusters. 
Our value (7.3 meV) is also in better agreement with the correlated model, although significantly different.
Rakvin and Dalal \cite{rakvin} and Hukuda \cite{hukuda} independently found a much higher activation energy (190 meV) from electronic spin resonance measurements, 
but their measurements cannot attribute it to a particular configuration.

\section{Discussion and conclusion}

The results presented above show very clearly that forbidden reflections are sensitive to proton configurations in hydrogen bonds.
We have presented above a method to analyse quantitatively their spectra and extract the relative probabilities of the configurations.
We find that the concentration of the polar and Slater configurations in KDP has a strong temperature dependence which follows the Arrhenius law.
The resulting activation energies for polar and Slater configurations are in a reasonable order of magnitude but are in reversed order compared to the usually accepted model. 
This suggests that the quantitative analysis of the spectra requires further development.

By comparison with the results obtained from the $006$ and $550$
forbidden reflections of RDP \cite{richter14}, 
whose crystal symmetry is isomorphic to KDP, 
the temperature effects are stronger in KDP. 
First, the phase transition from the para- to ferroelectric
phase provides a much stronger intensity jump at the $002$ reflection 
of KDP \cite{beutier14} than at the $006$ reflection of RDP.
Second, the energy spectra of the forbidden reflections 
are more complicated in KDP than in RDP and their variation with
temperature is also stronger. 
This spectroscopic difference is the result of a structural difference: 
while both materials have the same symmetry, 
Rb atoms are heavier than potassium atoms and the lattice
is possibly more rigid. 

Moreover, we evidenced in KDP the contributions of the
polar proton configurations, while in RDP only
the presence of the Slater configurations had been observed (in the $550$ reflection). 
This could be achieved by collecting a larger set of experimental data in KDP, 
\textit{i.e.}, two reflections at two azimuths each, 
using the same simple theoretical framework.
Moreover, measuring the spectra of the forbidden reflections 
at various azimuthal angles over a large temperature range
gives additional information, which is very important for fitting the spectra. 
Nevertheless, the method has a limited accuracy, 
which is illustrated by the quality of the fits in Fig. \ref{fits}. 
A number of simplifying assumptions contribute to this discrepancy: 
(a) In the calculation of the TMI term, 
we neglect the correlation of atomic displacements, 
which is equivalent to considering that only the resonant atoms vibrate; 
(b) proton distributions in a double-well potential
are considered as static configurations of defects 
and the structure amplitude is considered to be a sum of coherent contributions 
from the supercells with different kinds of defects; 
and (c) the various contributions to the resonant amplitude were calculated 
using the multiple scattering approach (as opposed to the finite-difference method) 
with a limited number of atoms involved in the multiple scattering model.
Nevertheless, this simple model provides a description of the main features 
of the forbidden reflections, their energy spectra and temperature dependence. 

As such, this study raises questions about the physics of hydrogen-bonded
materials, and provides a method to investigate them.
We should point out that while the potassium K edge provides a convenient resonance edge to apply the technique, 
the phosphorus K edge could be even more sensitive. 
Indeed, the potassium atoms are not directly related to the hydrogen bonds, 
while the phosphorus atoms are located at the centre of the hydrogen-bonded oxygen tetrahedra. 
The phosphorus K edge could possibly provide more accurate results and more detailed information about the proton configuration.
REXS at this edge would nevertheless imply experimental complications, due to the low energy (2.1455 keV).

Finally, we point out that the mechanism contributing to forbidden resonant reflections presented here, 
\textit{i.e.}, the proton disorder, is a realisation of the point-defect-induced scattering predicted in Ref. 
\onlinecite{dmitrienko2000}, 
which had not been evidenced experimentally prior to the studies of RDP and KDP. 
While it had been predicted as an effect of static disorder, here the protons are not static, 
but they are seen as such during the resonant x-ray scattering process.

\begin{acknowledgments}

The authors acknowledge the support of the XMaS staff
for the measurements performed at ESRF. 
XMaS is a mid-range facility supported by EPSRC. 
This work was supported by Grant No. RFBR 13-02-00760. 
The calculation have been performed using the
\textit{ab initio} total-energy and molecular-dynamics program VASP (Vienna
\textit{ab initio} simulation program) developed at the Institut für
Materialphysil of the Universitat Wien.
{ENO} is grateful to Professor B. A. Strukov for valuable discussions.

\end{acknowledgments}

\appendix

\section{Absorption spectra}
\label{absorption}

The spectra recorded in Bragg geometry are strongly modified by self-absorption \cite{fe3bo6}.
In order to properly analyse these spectra, the absorption spectrum must first be well characterized.

\begin{figure*}
    \includegraphics[width=\columnwidth]{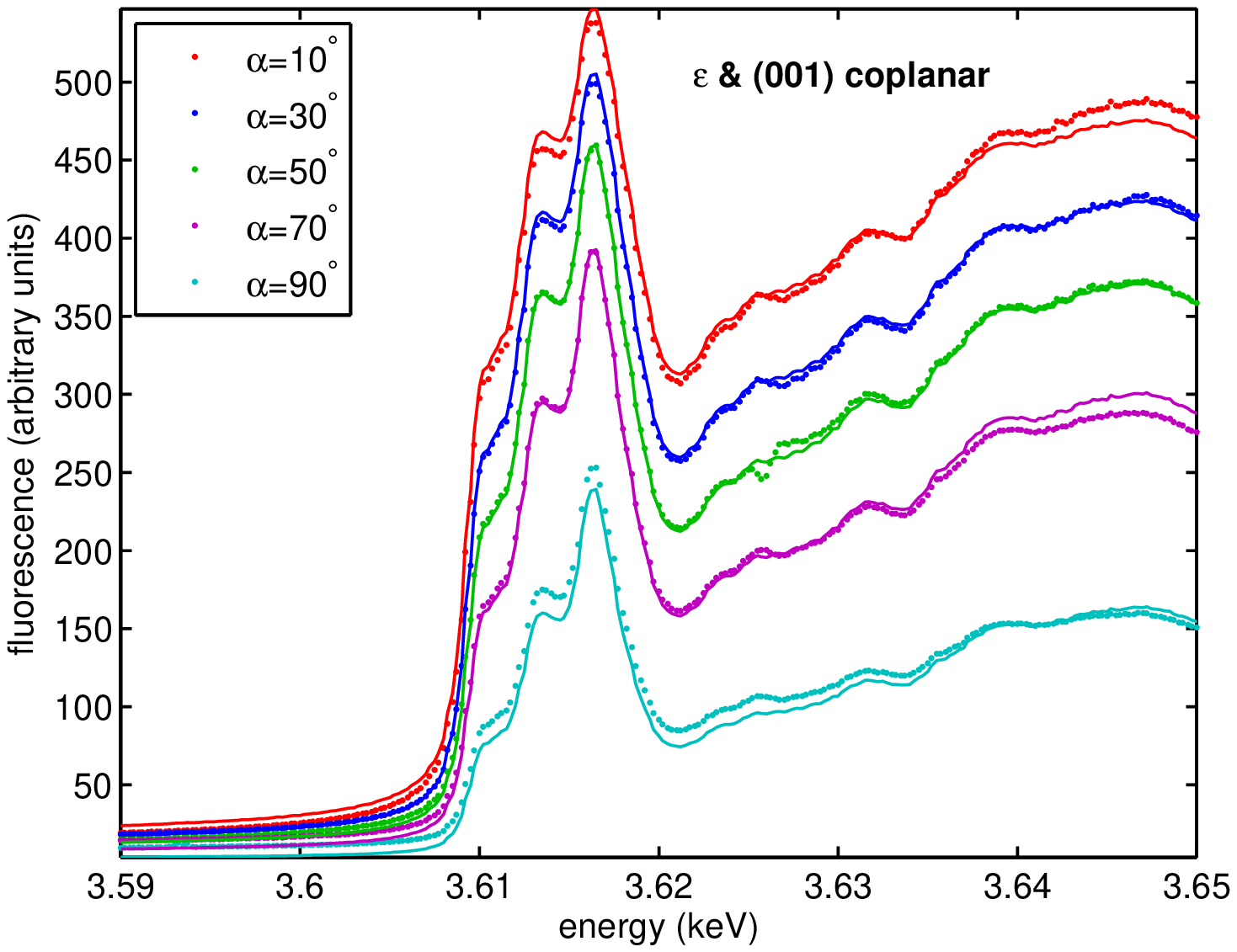}
    \includegraphics[width=\columnwidth]{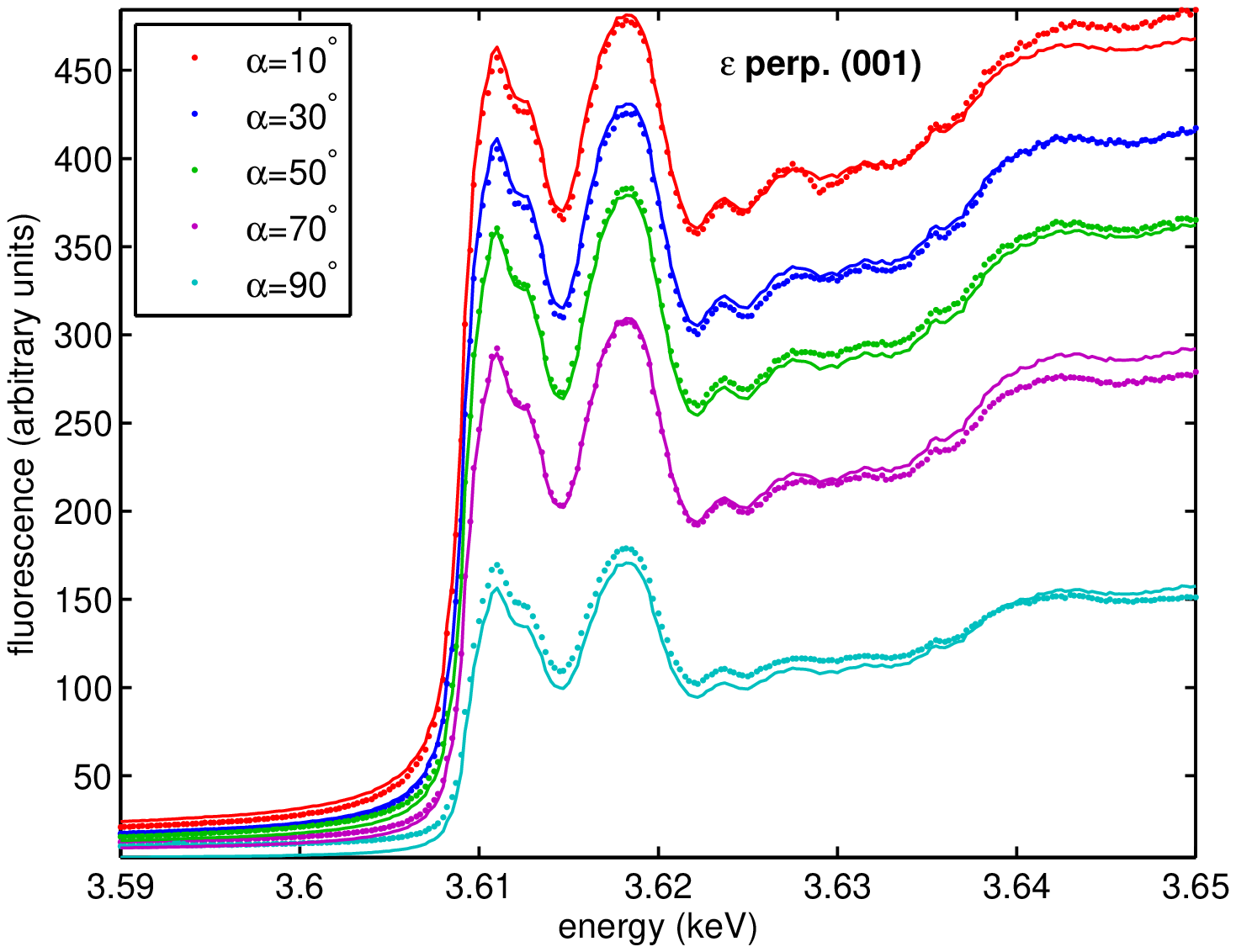}
    \caption{Fluorescence. 
    Left: Tetragonal axis coplanar with the polarisation of the incident beam. 
    Right: Tetragonal axis perpendicular to the polarisation of the incident beam.
    The data are recorded for various incident angles $\alpha$, with the detector at 100$^\circ$ from the incident beam. 
    In both panels, the dots show the data and the lines show the fits obtained with the absorption spectra shown in Fig. \ref{fig.mu}.}
    \label{fig.fluo}
\end{figure*}

The absorption cross section of noncubic crystals is anisotropic.
Tetragonal crystals, such as the paraelectric phase of KDP, display linear dichroism (in the electric dipole approximation) \cite{brouder}
and the absorption cross section can be decomposed into an isotropic part and an anisotropic (dichroic) part.
Though the polarisation vector is generally not an eigenstate of the optical system, we can make this approximation if the anisotropy is not too strong compared to the isotropic absorption.
Within this approximation, the linear absorption coefficient $\mu$ is proportional to the absorption cross-section.
According to Brouder's formalism \cite{brouder}, $\mu$ can be written as
\begin{equation}
    \mu(E,\eta) = \mu_{iso}(E)-\frac{3\cos^2\eta-1}{\sqrt{2}}\mu_{dic}(E)
\end{equation}
where $\eta$ is the angle between the beam polarisation and the tetragonal axis, $\mu_{iso}$ is the isotropic part, and $\mu_{dic}$ is the dichroic part.

\begin{figure}
    \includegraphics[width=\columnwidth]{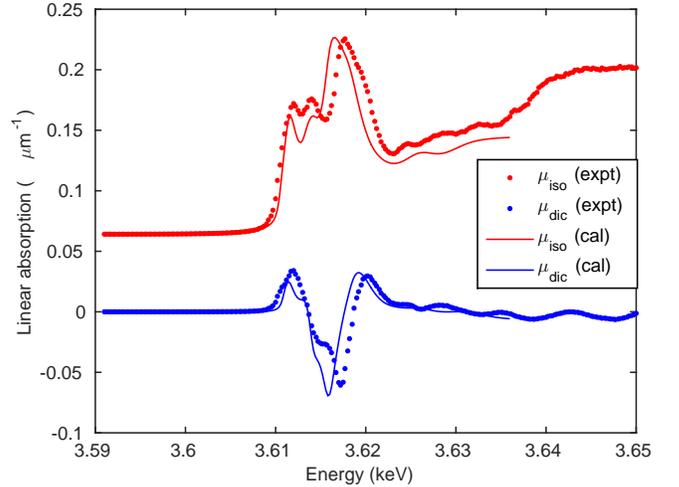}
    \caption{Isotropic and dichroic parts of the absorption coefficient, as determined from the fluorescence data (expt) and FDMNES calculations (cal).}
    \label{fig.mu}
\end{figure}

Two sets of fluorescence spectra with various incidence angles were recorded at beam line I16 of Diamond Light Source, at room temperature, 
the first one with the polarisation parallel to the tetragonal axis and the second one with the polarisation perpendicular to it. 
The fluorescence was measured in reflection geometry with the detector at 100$^\circ$ from the incident beam from a crystal with a (110) cut (Fig. \ref{fig.fluo}). 
The isotropic part and anisotropic parts of the absorption spectrum were extracted following a similar procedure to that detailed in 
Ref. \onlinecite{fe3bo6}. 
The results are shown in Fig. \ref{fig.mu} and reveal a rather strong linear dichroism.

The experimental absorption spectra obtained with this procedure were then
modelled with FDMNES \cite{bunau,fdmnes}. The calculations
were made with the finite difference method with a cluster of
7.5 \AA \ (145 atoms), and the simulated spectra were convoluted with the width of the excited state, described by an arctangent model.
Figure \ref{fig.mu} shows the isotropic and
anisotropic parts of the calculated absorption in comparison with
the experimental results.
The agreement between FDMNES calculations and experimental spectra is rather good for both spectra.

\section{Analysis of the ferroelectric phase}

\begin{figure}
    \includegraphics[width=\columnwidth]{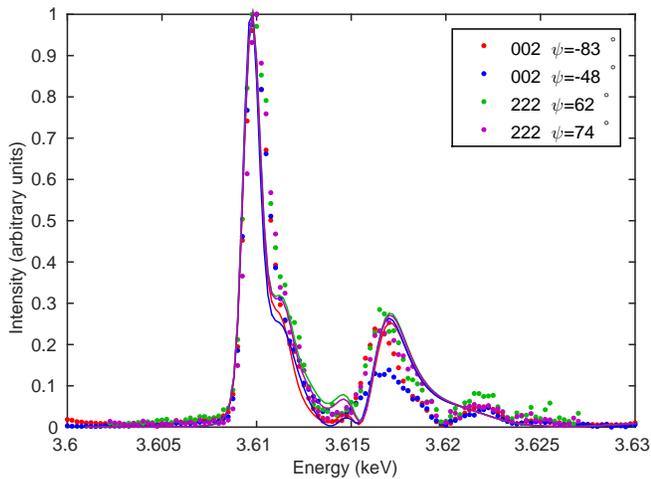}
    \caption{Energy spectra normalized to their maximum of the $002$ reflection at azimuths $-83^\circ$ and  $-48^\circ$,
      and of the $222$ reflection at azimuths $62^\circ$ and  $74^\circ$, at 80 $K$. 
      Dots: measurements; solid lines: calculations.}
    \label{lowT}
\end{figure}

In the ferroelectric phase, the resonant scattering factor is largely
dominated by the electric dipole-dipole (E1E1) contribution, whose
appearance explains the sudden changes of intensity and spectrum
across the phase transition \cite{beutier14}. 
It has no strong temperature dependence as long as it remains below the phase transition:
the lattice contraction has very little effect on the resonant
scattering if we consider only the static perfect structure \cite{oreshko12}. 
The variation of the forbidden reflections intensity 
observed below the phase transition in the experimental
data is ascribed to the rotation of orthorhombic domains. A
single E1E1 component contributes to the forbidden reflections, so
that the 002 and 222 reflections show nearly identical spectra,
independently of the azimuth \cite{beutier14}. 
Figure \ref{lowT} shows the comparison of the calculated and
experimental energy spectra of the $002$ and $222$ reflections
(each for two azimuthal angles). 
The calculated spectra are in good agreement with the experimental data. 
A small $hkl$ and azimuthal dependence of the calculated energy spectra can be noticed
between 3.612 and 3.616 keV and is ascribed to a minor contribution of the electric
dipole-quadrupole (E1E2) resonance.

\bibliography{journals.bib,kdp.bib,rexs.bib}

\end{document}